\documentclass[final,3p,times,authoryear]{elsarticle}

\usepackage{amssymb}
\usepackage{amsthm}
\usepackage{amsmath}
\usepackage{tabu} 
\usepackage[T1]{fontenc}
\usepackage{hyperref}
\usepackage{csquotes}
\usepackage{caption}
\usepackage{multirow}
\usepackage{multicol}
\usepackage{subfigure}

\graphicspath{{figures/}}

\newcommand\restr[2]{{\left.\kern-\nulldelimiterspace#1\right|_{#2}}}

\DeclareMathOperator*{\argmin}{arg\,min}

\newdefinition{definition}{Definition}

\journal{{}}

\begin{document}

\begin{frontmatter}

\title{Determination of an Optimal Control Strategy for Vaccine Administration in COVID-19 Pandemic Treatment}

\author[1]{Gustavo Barbosa Libotte}
\ead{gustavolibotte@iprj.uerj.br}
\author[2]{Fran S{\'e}rgio Lobato\corref{cor1}}
\ead{fslobato@ufu.br}
\cortext[cor1]{Corresponding author}
\author[3]{Gustavo Mendes Platt}
\ead{gmplatt@furg.br}
\author[1]{Ant{\^o}nio J. Silva Neto}
\ead{ajsneto@iprj.uerj.br}

\address[1]{Polytechnic Institute, Rio de Janeiro State University, Nova Friburgo, Brazil}

\address[2]{Chemical Engineering Faculty, Federal University of Uberl{\^{a}}ndia, Uberl{\^{a}}ndia, Brazil}

\address[3]{School of Chemistry and Food, Federal University of Rio Grande, Santo Ant\^{o}nio da Patrulha, Brazil}

\begin{abstract}
For decades mathematical models have been used to predict the behavior of physical and biological systems, as well as to define strategies aiming at the minimization of the effects regarding different types of diseases. In the present days, the development of mathematical models to simulate the dynamic behavior of the novel coronavirus disease (COVID-19) is considered an important theme due to the quantity of infected people worldwide. In this work, the aim is to determine an optimal control strategy for vaccine administration in COVID-19 pandemic treatment considering real data from China. For this purpose, an inverse problem is formulated and solved in order to determine the parameters of the compartmental SIR (Susceptible-Infectious-Removed) model. To solve such inverse problem, the Differential Evolution (DE) algorithm is employed. After this step, two optimal control problems (mono- and multi-objective) to determine the optimal strategy for vaccine administration in COVID-19 pandemic treatment are proposed. The first consists of minimizing the quantity of infected individuals during the treatment. The second considers minimizing together the quantity of infected individuals and the prescribed vaccine concentration during the treatment, i.e., a multi-objective optimal control problem. The solution for both optimal control problems is obtained using DE and Multi-Objective Differential Evolution (MODE) algorithms, respectively. The results regarding the proposed multi-objective optimal control problem provides a set of evidences from which an optimal strategy for vaccine administration can be chosen, according to a given criterion.
\end{abstract}




\end{frontmatter}


\section{Introduction}

In the last decades, countless mathematical models used to evaluate the spread and control of infectious diseases have been proposed. These models are very important in different fields, such as policy making, emergency planning and risk assessment, definition of control-programs, and promotion of the improvement of various health-economic aspects \citep{AlSheikh2012}. In general, such models aim to describe a state of infection (susceptible and infected) and a process of infection (the transition between these states) by using compartmental relations, i.e., the population is divided into compartments by taking assumptions about the nature and time rate of transfer from one compartment to another \citep{Trawicki2017,BlackwoodChilds2018}. One can cite several studies using models for measles vaccination \citep{BauchSzuszGarrison2009,Widyaningsih2018}, HIV/AIDS \citep{MukandavireChiyakaGariraMusuka2009}, tuberculosis \citep{BowongKurths2010}, dengue \citep{Weiss2013}, pertussis epidemiology \citep{Pesco2014}, among others.

Recently, for the past five to six months, the world has been experiencing the dissemination of a new virus, referred to as COVID-19 (Coronavirus disease 2019). COVID-19 is an infectious disease emerged from China in November 2019, that has rapidly spread around in many other countries worldwide \citep{GorbalenyaGulyaeva2020,WorldHealthOrganization}. The common symptoms are severe respiratory illness, fever, cough, and myalgia or fatigue, especially at the onset of illness \citep{HuangWangLiRenZhaoHuZhangFanXuGu2020}. The transmission may happen person-to-person, through direct contact or droplets \citep{ChanYuanKokToChu2020,LiGuanWu2020,RiouAlthaus2020}.

Since the COVID-19 outbreak in Wuhan City in November of 2019, various computational model-based predictions have been proposed and studied. \cite{LinZhaoGaoLouYangMusaWangCaiWangYangHe2019} proposed a Susceptible-Exposed-Infectious-Removed (SEIR) model for the COVID-19 outbreak in Wuhan. These authors considered some essential elements including individual behavioral response, governmental actions, zoonotic transmission and emigration of a large proportion of the population in a short time period. \cite{BenvenutoGiovanettiVassalloAngelettiCiccozzi2020} proposed the Auto Regressive Integrated Moving Average (ARIMA) model to predict the spread, prevalence and incidence of COVID-2019. \cite{RodaVarugheseHanLi2020} used a Susceptible-Infectious-Removed (SIR) model to predict the COVID-19 epidemic in Wuhan after the lockdown and quarantine. In such study, the authors demonstrate that non-identifiability in model calibrations using the confirmed-case data is the main reason for wide variations in the results. \cite{PremLiuRussellKucharskiEggo2020} proposed a SEIR model to simulate the spread of COVID-19 in Wuhan city. In this model, all demographic changes in the population (births, deaths and ageing) were ignored. The simulations showed that control measures aimed at reducing social mixing in the population can be effective in reducing the magnitude and delaying the peak of the COVID-19 outbreak.

In order to evaluate the global stability and equilibrium point of these models, \cite{LiMuldowney1995} studied a SEIR model with nonlinear incidence rates in epidemiology, in terms  of global stability of endemic equilibrium. \cite{AlSheikh2012} evaluated a SEIR epidemic model with limited resources for treating infected people. For this purpose, the existence and stability of disease-free and endemic equilibrium were investigated. \cite{LiCui2013} studied a SEIR model with vaccination strategy that incorporates distinct incidence rates for exposed and infected populations. These authors proved the global asymptotical stable results of the disease-free equilibrium. \cite{SinghSrivastavaArora2017} developed a simple and effective mathematical model for transmission of infectious diseases by taking into consideration the human immunity. This model was evaluated in terms of local stability of both disease free equilibrium and disease endemic equilibrium. \cite{Widyaningsih2018} proposed a SEIR model with immigration and determined the system equilibrium conditions. \cite{KimByunJung2019} developed a Coxian-distributed SEIR model considering an empirical incubation period, and a stability analysis was also performed.

In order to reduce the dissemination of COVID-19 worldwide, various procedures have been adopted. As mentioned by \cite{PanZhaiYanbingDingXiaWuJunkeLongYanjunZhongYimingLi2020} and \cite{WeiZhengLeiWuVermaLiuWeiBiHuHan2020}, quarantine and isolation (social-distancing) can effectively reduce the spread of COVID-19. In addition, wearing masks, washing hands and disinfecting surfaces contribute to reducing the risk of infection. According to the U.S. Food and Drug Administration, there are no specific therapies to COVID-19 treatment. However, treatments including antiviral agents, chloroquine and hydroxychloroquine, corticosteroids, antibodies, convalescent plasma transfusion and radiotherapy are being studied \citep{WangHuHu2020}.

As alternative to these treatments, the use of drug administration (vaccine) arises as an interesting alternative to face this pandemic. It must be emphasized that there is currently no vaccine to COVID-19, but there is a huge effort to develop a vaccine in a record time, which justifies the present study \citep{Lurie2020}. Mathematically, the determination of optimal protocol for vaccine administration characterizes an Optimal Control Problem (OCP). This particular optimization problem consists in the determination of control variable profiles that minimize (or maximize) a given performance index \citep{BrysonHo1975,Biegleretal2002}. In order to solve this problem, several numerical methods have been proposed \citep{BrysonHo1975,Feehery1998,Lobato2004,LobatoSilverioSteffen2016}. These methods are classified according to three broad categories: direct optimization methods, Pontryagin's Maximum Principle (PMP) based methods and HJB-based (Hamilton-Jacob-Bellman) methods. The direct approach is the most traditional strategy considered to solve an OCP, due to its simplicity. In this approach, the original problem is transformed into a finite dimensional optimization problem through the parametrization of control or parametrization of control and state variables \citep{Feehery1998}.

From an epidemiological point of view, \cite{NeilanLenhart2010} proposed an optimal control problem to determine a vaccination strategy over a specific period of time so as to minimize a cost function. In this work, the propagation of a disease is controlled by a limited number of vaccines, while minimizing a percentage of the overall number of dead people by infection, and a cost associated with vaccination. \cite{BiswasPaivaPinho2014} studied different mathematical formulations for an optimal control problem
considering a Susceptible-Exposed-Infectious-Removed model. For this purpose, these authors evaluated the solution of such problems when mixed state control constraints are used to impose upper bounds on the available vaccines at each instant of time. In addition, the possibility of imposing upper bounds on the number of susceptible individuals with and without limitations on the number of vaccines available were analyzed. The optimal control theory was applied to obtain optimal vaccination schedules and control strategies for the epidemic model of human infectious diseases.

In this work, the objective is to determine an optimal control strategy for vaccine administration in COVID-19 pandemic treatment considering real data from China. In order to determine the parameters that characterize the proposed mathematical model (based on the compartmental SIR model), an inverse problem is formulated and solved considering the Differential Evolution (DE) algorithm \citep{StornPrice1995,StornPriceLampinen}. After this step, two optimal control problems (mono- and multi-objective) used to determine the optimal strategy for vaccine administration in COVID-19 pandemic treatment are proposed. The mono-objective optimal control problem considers minimizing the quantity of infected individuals during the treatment. On the other hand, the multi-objective optimal control problem considers minimizing together the quantity of infected individuals and the prescribed vaccine concentration during the treatment. To solve each problem, DE and Multi-Objective Differential Evolution (MODE) algorithms \citep{LobatoSteffenJr} are employed, respectively.

This work is organized as follows. Section~\ref{sec:mat_model} presents the description of the mathematical model considered to represent the evolution of COVID-19 pandemic. In Section~\ref{sec:form_ocp}, the general aspects regarding the formulation and solution of an OCP is presented. A brief review on DE and its extension to deal with multi-criteria optimization is presented in Section~\ref{sec:de_descript}. In Section~\ref{sec:methodology}, the proposed methodology is presented and discussed. The results obtained using such methodology are presented in Section~\ref{eq:results}. Finally, the conclusions are outlined in Section~\ref{eq:conclusions}.

\section{Mathematical Modeling in Epidemiology} \label{sec:mat_model}

In the specialized literature, various compartmental models used to represent the evolution of an epidemic can be found \citep{Forgoston2013,Pesco2014,Shaman2014,Cooper2016,ShumailaAzametal}. The study of these models is very important to understand the epidemic spreading mechanisms and, consequently, to investigate the transmission dynamics in population \citep{Forgoston2013}. As mentioned by \cite{KeelingRohani2011}, these compartmental models can be divided into two groups: \textit{i}) population-based models and \textit{ii}) agent-based or individual-based models. In turn, the first one can be subdivided into deterministic or stochastic (considering continuous time, ordinary differential equations, partial differential equations, delay differential equations or integro-differential equations) or discrete time (represented by difference equations). The second class can be subdivided into usually stochastic and usually discrete time.

In the context of population-based models, the deterministic modeling can be represented, in general, by the interaction among susceptible (denoted by $S$ --- an individual which is not yet infected by the disease pathogen), exposed (denoted by $E$ --- an individual in the incubation period after being infected by the disease pathogen, and with no visible clinical signs), infected/infectious (denoted by $I$ --- an individual that can infect others) and, recovered individuals (denoted by $R$ --- an individual who survived after being infected but is no longer infectious and has developed a natural immunity to the disease pathogen). Considering a population of size $ N $, and based on the disease nature and on the spreading pattern, the compartmental models can be represented as \citep{KeelingRohani2011,Hethcote2000}:

\begin{itemize}
  \item Susceptible-Infected (SI): population described by groups of Susceptible and Infected;
  \item Susceptible-Infected-Removed (SIR): population described by groups of Susceptible, Infected and Recovered;
  \item Susceptible-Infectious-Susceptible (SIS): population also described by groups of Susceptible and Infected. In this particular case, recovering from some pathologies do not guarantee lasting immunity. Thus, individuals may become susceptible again;
  \item Susceptible-Exposed-Infectious-Removed (SEIR): population described by groups of Susceptible Exposed, Infected and Recovered.
\end{itemize}

It is important to mention that in all these models, terms associated with birth, mortality and vaccination rate can be added. In addition, according to
\cite{KeelingRohani2011} and \cite{Hethcote2000}, these models can include: \textit{i}) time-dependent parameters to represent the effects of seasonality; \textit{ii}) additional compartments to model vaccinated and
asymptomatic individuals, and different stages of disease progression; \textit{iii}) multiple groups to model heterogeneity, age, spatial structure or host species; \textit{iv}) human demographics parameters, for diseases where the time frame of the disease dynamics is comparable to that of human demographics. Human demographics can be modeled by adopting constant immigration rate, constant \textit{per capita} birth and death rates, density-dependent death rate or disease-induced death rate. Thus, the final model is dependent on assumptions taken during the formulation of the  problem.

\begin{figure}[!htbp]
\centering
\includegraphics[scale=1]{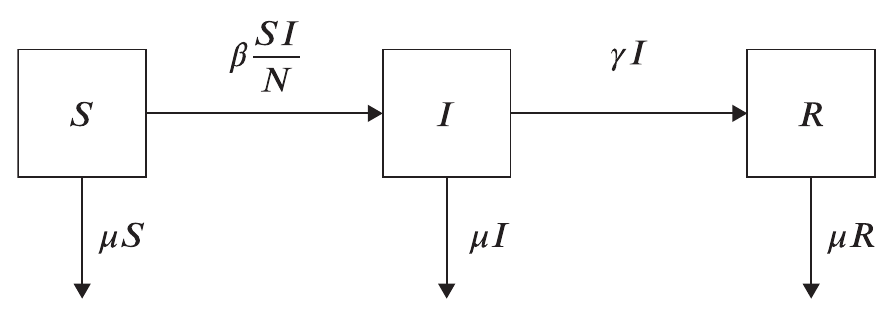}
\caption{Compartments in the SIR model \citep{KeelingRohani2011}.} \label{Fig1}
\end{figure}

In this work, the SIR model is adopted, in order to describe the dynamic behavior of COVID-19 epidemic in China. The choice of this model is due to the study conducted by \cite{RodaVarugheseHanLi2020}. These authors demonstrated that the SIR model performs more adequately than the SEIR model in representing the information related to confirmed case data. For this reason, the SIR model will be adopted here. The schematic representation of this model is presented in Fig.~\ref{Fig1}.

Mathematically, this model has the following characteristics:

\begin{itemize}
\item An individual is susceptible to an infection and the disease can be transmitted from any infected individual to any susceptible individual. Each susceptible individual is given by the following relation:
\begin{equation}
\label{Eq01}
\frac{dS}{dt}=-\beta\frac{SI}{N}-\mu S, \quad S(0)=S_0
\end{equation}
where $t$ is the time, $\beta$ and $\mu$ represents the probability of transmission by contact and \textit{per capita} removal rate, respectively. In turn, $S_0$ is the initial condition for the susceptible population.
%
%
\item Any infected individual may transmit the disease to a susceptible one according to the following relation:
\begin{equation}
\label{Eq03}
\frac{dI}{dt}=\beta\frac{SI}{N}-(\gamma+\mu) I, \quad I(0)=I_0
\end{equation}
where $\gamma$ denotes the \textit{per capita} recovery rate. $I_0$ is the initial condition for the infected population.
\item Once an individual has been moved from Infected to Recovered, it is assumed that it is not possible to be infected again. This condition is described by:
\begin{equation}
\label{Eq04}
\frac{dR}{dt}=\gamma I-\mu R, \quad R(0)=R_0
\end{equation}
where $R_0$ is the initial condition for the recovered population.
\end{itemize}

It is important to emphasize that the population size ($N$) along time $t$ is defined as $N ( t ) = S ( t ) + I ( t ) + R ( t ) $. In practice, the model parameters must be determined to represent a particular epidemic. For this purpose, it is necessary to formulate and to solve an inverse problem. In the section that describes the methodologies adopted in this work, more details on the formulation and solution of this problem is presented.

\section{Formulation of the Optimal Control Problem} \label{sec:form_ocp}

Mathematically, an OCP can be formulated as follows \citep{BrysonHo1975, Feehery1998, Lobato2004}. Initially, let
\begin{equation}
\label{EQ1c} J = \Psi \left( z \left( t_f \right), \; t_f \right) + \int\limits_{t_0}^{t_f} {L \left(z, \; u, \; t \right) \; \mathrm{d}t}
\end{equation}

\noindent where $z$ is the vector of state variables, and $u$ is the vector of control variables. $\Psi$ and the integration of $L$ over a period of time [$t_0$ $t_f$] are the first and second terms of the performance index, respectively. The minimization problem is given by
\begin{align} \label{eq:OCPdef}
\begin{split}
\argmin\limits_{u(t), \; t_f } \; &J\\
\textrm{Subject to} \; &f(\dot z,z,u,t) = 0\\
\hphantom{\textrm{Subject to}} &g(z,u,t) \le 0\\
\hphantom{\textrm{Subject to}} &p(u,t) \le 0\\
\hphantom{\textrm{Subject to}} & \left. {q(z,u,t)} \right|_{t \; = \; t_f }  = 0
\end{split}
\end{align}

\noindent with consistent initial conditions given by
\begin{equation}
\label{EQ6c} \varphi (\dot z(t_0 ),z(t_0 ),u(t_0 ),t_0 ) = 0
\end{equation}

\noindent where $J(.)$, $L(.)$, $\Psi(.) \rightarrow \mathrm{I\!R}$; $f(.), \varphi(.) \rightarrow \mathrm{I\!R}^{m_z}$; $z \in \mathrm{I\!R}^{m_z}$;  $u \in \mathrm{I\!R}^{m_u}$; $g \in \mathrm{I\!R}^{m_g}$; $p \in \mathrm{I\!R}^{m_p}$ and; $q \in \mathrm{I\!R}^{m_q}$.
 
According to the optimal control theory \citep{BrysonHo1975,Feehery1998}, the solution of the OCP, whose problem is defined by Eqs.~(\ref{eq:OCPdef}) and (\ref{EQ6c}), is satisfied by the co-state equations and the stationary condition given, respectively, by
\begin{equation}
\label{EQ7c} \dot \lambda ^{\mathrm{T}}  \equiv  - \frac{{\partial H}}{{\partial z}}{\rm{,  \;\;\;\;\;\;\;    }}\lambda (t_f ) = \left. {\frac{{\partial \Psi }}{{\partial z}}} \right|_{t = t_f }
\end{equation}
\begin{equation}
\label{EQ8c} \frac{{\partial H}}{{\partial u}} = 0
\end{equation}

\noindent where $H$ is the Hamiltonian function defined by
\begin{equation}
\label{EQ9c} H \equiv L + \lambda ^{\mathrm{T}} f
\end{equation}

This system of equations is known as the Euler-Lagrange equations (optimality conditions), which are characterized as Boundary Value Problems (BVPs). Thus, to solve this model, an appropriated methodology must be used, as for example, the Shooting Method or the Collocation Method \citep{BrysonHo1975}. As mentioned by \cite{BrysonHo1975} and \cite{Feehery1998}, the main difficulties associated with OCPs are the following: the existence of end-point conditions (or region constraints) implies multipliers and associated complementary conditions that significantly increase the complexity of solving the BVP using an indirect method; the existence of constraints involving the state variables and the application of slack variables method may introduce differential algebraic equations of higher index; the Lagrange multipliers may be very sensitive to the initial conditions.

\section{Differential Evolution and Multi-objective Optimization Differential Evolution - A Brief Description} \label{sec:de_descript}

\subsection{Differential Evolution} \label{sec:DE}

Differential Evolution is a powerful optimization technique to solve mono-objective optimization problems, proposed by \cite{StornPrice1995}. This evolutionary strategy differs from other population-based algorithms in the schemes considered to generate a new candidate to solution of the optimization problem \citep{StornPrice1995,StornPriceLampinen,LobatoSteffenJr}. The population evolution proposed by DE follows three fundamental steps: mutation, crossover and selection. The optimization process starts by creating a vector containing $ NP $ individuals, called initial population, which are randomly distributed over the entire search space. During $ G_{\text{max}} $ generations, each of the individuals that constitute the current population are subject to the procedures performed by the genetic operators of the algorithm.

In the first step, the mutation operator creates a trial vector by adding the balanced difference between two individuals to a third member of the population, by $ v_{j}^{\left(G+1\right)} = x_{{\kappa_{1}}}^{\left(G\right)} + F\left(x_{{\kappa_{2}}}^{\left(G\right)} - x_{{\kappa_{3}}}^{\left(G\right)}\right) $, where $j = 1, \dots, \; NP$. The parameter $ F $ represents the scale factor, which controls the contribution added by the vector difference, such that $ F \in \left[0,\;2\right] $. In turn, \cite{StornPrice1995} proposed various mutation schemes for the generation of trial vectors (candidate solutions) by combining the vectors that are randomly chosen from the current population, such as:
\begin{itemize}
  \item rand/1: $x = x_{\kappa_1}  + F\left( {x_{\kappa_2}  - x_{\kappa_3} } \right)$
  \item rand/2: $x = x_{\kappa_1}  + F\left( {x_{\kappa_2}  - x_{\kappa_3}  + x_{\kappa_4}  - x_{\kappa_5} }\right)$
  \item best/1: $x = x_{\mathrm{best}}  + F\left( {x_{\kappa_2}  - x_{\kappa_3} } \right)$
  \item best/2: $x = x_{\mathrm{best}}  + F\left( {x_{\kappa_2}  - x_{\kappa_3}  + x_{\kappa_4}  - x_{\kappa_5} } \right)$
  \item rand/best/1: $x = x_{\kappa_1}  + F\left( {x_{\mathrm{best}}  - x_{\kappa_1}  + x_{\kappa_1}  - x_{\kappa_2} } \right)$
  \item rand/best/2: $x = x_{\kappa_1}  + F\left( {x_{\mathrm{best}}  - x_{\kappa_1} } \right) + F\left({x_{\kappa_1}  - x_{\kappa_2}  + x_{\kappa_3}  - x_{\kappa_4} } \right)$
\end{itemize}

The second step of the algorithm is the crossover procedure. This genetic operator creates new candidates by combining the attributes of the individuals of the original population with those resulting in the mutation step. The vector $ u_{jk}^{\left(G+1\right)} = v_{jk}^{\left(G+1\right)} $ if $ randb\left(k\right) \leq CR $ or $ k = rnbr\left( j \right) $. Otherwise, $ u_{jk}^{\left(G+1\right)} = x_{jk}^{\left(G\right)} $, such as $k = 1, \dots, \; d$, where $ d $ denotes the dimension of the problem and $randb\left(k\right) \in \left[0, \; 1\right]$ is a random real number with uniform distribution. The choice of the attributes of a given individual is defined by the crossover probability, represented by $ CR $, such that $ CR \in \left[0,\;1\right] $ is a constant parameter defined by the user. In turn, $rnbr \left( j \right) \in \left[1,\;d\right]$ is a randomly chosen index.

After the generation of the trial vector by the steps of mutation and crossover, the evolution of the best individuals is defined according to a greedy strategy, during the selection step. \cite{StornPriceLampinen} have defined some simple rules for choosing the key parameters of DE for general applications. Typically, one might choose \textit{NP} in the range from 5 to 10 times the dimension ($ d $) of the problem. In the case of \textit{F}, it is suggested taking a value ranging between 0.4 and 1.0. Initially, \textit{F} = 0.5 may be a good choice. In the case of premature convergence, \textit{F} and \textit{NP} may be increased.

\subsection{Multi-objective Optimization Differential Evolution} \label{sec:MODE}

The multi-objective optimization problem (MOP) is an extension of the mono-objective optimization problem. Due to the conflict between the objectives, there is no single point capable of optimizing all functions simultaneously. Instead, the best solutions that can be obtained are called optimal Pareto solutions, which form the Pareto curve \citep{Deb}. The notion of optimality in a MOP is different from the one regarding optimization problems with a single objective. The most common idea about multi-objective optimization found in the literature was originally proposed by \cite{Edgeworth}, and further generalized by \cite{Pareto}. One solution is said to be dominant over another, if it is not worse in any of the objectives, and if it is strictly better in at least one of the objectives. As an optimal Pareto solution dominates any other feasible point in the search space, all of these solutions are considered better than any other. Therefore, multi-objective optimization consists of finding a set of points that represents the best balance in relation to minimizing all objectives simultaneously, that is, a collection of solutions that relates the objectives, which are in conflict with each other, in most cases.

Let $ \boldsymbol{F} \left( \boldsymbol{x} \right) = \left( F_{1} \left( \boldsymbol{x} \right) , \; \dots, \; F_{m} \left( \boldsymbol{x} \right) \right)^{\mathrm{T}} $ be the \textit{objective vector} such that $ F_{k} : P \rightarrow \mathrm{I\!R} $, for $ k = 1, \; \dots, \; m $, where $ \boldsymbol{x} \in P $ is called \textit{decision vector} and its entries are called \textit{decision variables} and $m$ is the number of objective functions. Mathematically, a MOP is defined as \citep{Deb, Lobato2008}:
\begin{align*}
\begin{split}
\min \; &\boldsymbol{F} \left( \boldsymbol{x} \right)\\
\textrm{Subject to} \; &G_{i} \left( \boldsymbol{x} \right) \leq 0\\
\hphantom{\textrm{Subject to}} &H_{j} \left( \boldsymbol{x} \right) = 0\\
\hphantom{\textrm{Subject to}} &\boldsymbol{x}_{\mathrm{inf}} \leq \boldsymbol{x} \leq \boldsymbol{x}_{\mathrm{sup}}
\end{split}
\end{align*}
where $G$ is the vector of inequality constraints and $H$ is the vector of equality constraints.

Due to the favorable outcome of DE in solving mono-objective optimization problems, for different fields of science and engineering, \cite{LobatoSteffenJr} proposed the Multi-Objective Differential Evolution (MODE) algorithm to solve multi-objective optimization problems. Basically, this evolutionary strategy differs from other algorithms by the incorporation of two operators to the original DE algorithm, the mechanisms of rank ordering \citep{Deb,Zitzler}, and exploration of the neighborhood for potential solution candidates \citep{Hu}. A brief description of the algorithm is presented next.

At first, an initial population of size $NP$ is randomly generated, and all objectives are evaluated. All dominated solutions are removed from the population by using the operator \textit{Fast Non-Dominated Sorting} \citep{Deb}. This procedure is repeated until each candidate vector becomes a member of a front. Three parents generated by using DE algorithm are selected at random in the population. Then, an offspring is generated from these parents (this process continues until $NP$ children are generated). Starting from population $P_{1}$ of size $2 NP$, neighbors are generated to each one of the individuals of the population. These neighbors are classified according to the dominance criterion, and only the non-dominated neighbors ($P_{2}$) are put together with $P_{1}$, in order to form $P_{3}$. The population $P_{3}$ is then classified according to the dominance criterion. If the number of individuals of the population $P_{3}$ is larger than a predefined number, the population is truncated according to the \textit{Crowding Distance} \citep{Deb} criterion. This metric describes the density of candidate solutions surrounding an arbitrary vector. A complete description of MODE is presented by \cite{LobatoSteffenJr}.

\section{Methodology} \label{sec:methodology}

\subsection{Inverse Problem}

As mentioned earlier, the first objective of this work is to determine the parameters of the SIR model adopted to predict the evolution of COVID-19 epidemic considering experimental data from China. In this case, it is necessary to formulate and to solve an inverse problem. It arises from the requirement of determining parameters of theoretical models in such a way that it can be employed to simulate the behavior of the system for different operating conditions. Basically, the estimation procedure consists of obtaining the model parameters by the minimization of the difference between calculated and experimental values.

In this work, it is assumed that, since the outbreak persists for a relatively short period of time, the rate of births and deaths by natural cases or other reasons of the population is insignificant. Thus, we take $ \mu = 0 $, since there are probably few births/deaths in the corresponding period. 
We are interested in the determination of the following parameters of the SIR model: $\beta$, $\gamma$ and $I_0$. It is important to mention that $I_0$ is used to define the initial condition of all dependent variables of the model. Let
\begin{equation}
\mathcal{F} \equiv \sum\limits_{i = 1}^M {\frac{{\left( {I_i^{\exp }  - I_i^{{\rm{sim}}} } \right)^2 }}{{\left( {\max \left( {I^{\exp } } \right)} \right)^2 }}}
\end{equation}

\noindent Mathematically, the inverse problem is formulated as
\begin{equation}
\label{EQInv01} \argmin\limits_{\beta, \; \gamma, \; I_0 } \; \mathcal{F}
\end{equation}

\noindent subject to Eqs.~(\ref{Eq01}) -- (\ref{Eq04}), where $I_i^{\exp}$ and $I_i^{\rm{sim}}$ are the experimental and simulated infected population, respectively, and $M$ represents the total number of experimental data available. In this case, the SIR model must be simulated considering the parameters calculated by DE, in order to obtain the number of infected people estimated by the model and, consequently, the value of the objective function ($\mathcal{F}$). As the number of measured data, $M$, is usually much larger than the number of parameters to be estimated, the inverse problem is formulated as a finite dimensional optimization problem in which we aim at minimizing $\mathcal{F}$ \citep{MouraNetoSilvaNeto}.

\subsection{Optimal Control Problem}

\begin{figure}[!htbp]
\centering
\includegraphics[scale=1]{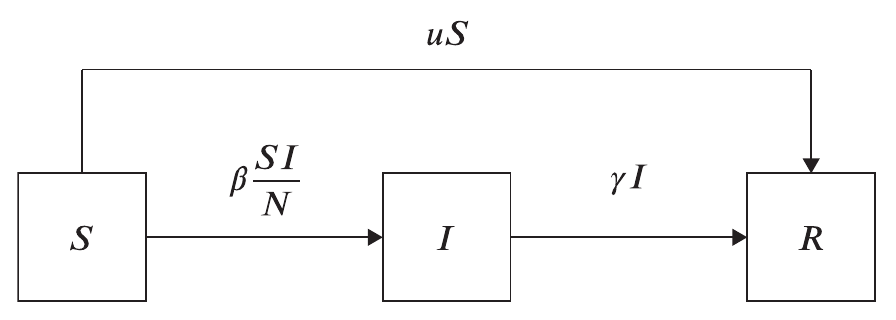}
\caption{Compartments in the SIR model with vaccination.} \label{Fig1vacina}
\end{figure}

In order to formulate both OCPs, the parameters estimated considering the proposed inverse problem are used. As proposed by \cite{NeilanLenhart2010} and \cite{BiswasPaivaPinho2014}, a new variable $W$, which denotes the number of vaccines used, is introduced in order to determine the optimal control strategy for vaccine administration. For this purpose, the total amount of vaccines available during the whole period of time is proportional to $u S$. Physically, $u$ represents the portion of susceptible individuals being vaccinated per unit of time \citep{BiswasPaivaPinho2014}. It is important to mention that $u$ acts as the control variable of such system. If $u$ is equal to zero there is no vaccination, and $u$ equals to one indicates that all susceptible population is vaccinated. A schematic diagram of the disease transmission among the individuals for the SIR model with vaccination is shown in Fig.~\ref{Fig1vacina}.

Mathematically, the SIR model considering the presence of control is written as:
\begin{equation}
\label{Eq01control}
\frac{dS}{dt}=-\beta\frac{SI}{N}-uS, \quad S(0)=S_0
\end{equation}
%
%
\begin{equation}
\label{Eq03control}
\frac{dI}{dt}=\beta\frac{SI}{N}-\gamma I, \quad I(0)=I_0
\end{equation}
\begin{equation}
\label{Eq04control}
\frac{dR}{dt}=\gamma I, \quad R(0)=R_0
\end{equation}
\begin{equation}
\label{Eq05control}
\frac{dW}{dt}=u S, \quad W(0)=W_0
\end{equation}

\noindent where $W_0$ is the initial condition for the total amount of vaccines. It is important to emphasize that the population size ($N$) after the inclusion of this new variable $W$ along the time $t$ is defined as $N ( t ) = S ( t ) + I ( t ) + R ( t ) + W ( t ) $.

The first formulation aims to determine the optimal vaccine administration ($u$) to minimize the infected population, represented by $\Omega_1$. Thus, let
\begin{equation} \label{eq:num_infected}
\Omega_1 \equiv \int_{t_0}^{t_f} {I \; \mathrm{d}t}
\end{equation}

\noindent The OCP is defined as
\begin{equation}
\label{EQOCPeq01} 
\argmin_u \; \Omega_1
\end{equation}

\noindent subject to Eqs.~(\ref{Eq01control}) -- (\ref{Eq05control}) and $ u_{\textrm{min}} \leq u \leq u_{\textrm{max}} $, where $t_0$ and $t_f$ represents the initial and the final time, respectively, and $u_{\textrm{min}}$ and $u_{\textrm{max}}$ are the lower and upper bounds for the control variable, respectively.

The second formulation considers two objectives, i.e., the determination of the optimal vaccine administration, in order to minimize the number of infected individuals and, at the same time, to minimize the number of vaccines needed. The total number of vaccines can be determined by
\begin{equation} \label{eq:num_vaccines}
\Omega_2 \equiv \int_{t_0}^{t_f} {u \; \mathrm{d}t}
\end{equation}

\noindent whereas the number of infected people is given by Eq.~(\ref{eq:num_infected}). Thus, the multi-objective optimization problem is formulated as

\begin{equation}
\label{EQOCPeq02}
\argmin_u \; \left( \Omega_1, \; \Omega_2 \right)
\end{equation}

\noindent subject to Eqs.~(\ref{Eq01control}) -- (\ref{Eq05control}) and $ u_{\textrm{min}} \leq u \leq u_{\textrm{max}} $. In both problems, the control variable $u$ must be discretized. In this context, the approach proposed consists on transforming the original OCP into a nonlinear optimization problem. For this purpose, let the time interval $ \left[ 0, \; t_f \right] $ be discretized using $ N_{\textrm{elem}} $ time nodes, with each node denoted by $ t_i $, where $ i = 0, \; \hdots, \; N_{\textrm{elem}} - 1 $, such that $ t_{0} \leq t_{i} \leq t_{f} $. For each of the $ N_{\textrm{elem}} - 1 $ subintervals of time, given by $ \left[ t_i, \; t_{i+1} \right] $, the control variable is considered constant by parts, that is, $ u \left( t \right) = u_{i} $ for $ t_{i} \leq t < t_{i + 1} $, where $ u_{\textrm{min}} \leq u_{i} \leq u_{\textrm{max}} $.

In order to obtain an optimal control strategy for vaccine administration, that can be used in medical practice, we consider the bang-bang control which consists of a binary feedback control that turns either \enquote{on} (in our case, when $u = u_{\textrm{max}} = 1$) or \enquote{off} (when $u = u_{\textrm{min}} = 0$) at different time points, determined by the system feedback. In this case, as the control strategy $u$ is constant by parts, the proposed optimal control problem has $ N_{\textrm{elem}} - 2 $ unknown parameters, since the control variable at the start and end times are known. The resulting nonlinear optimization problems are solved by using the DE, in the case of the mono-objective problem, given by Eq.~(\ref{EQOCPeq01}), and MODE, for the multi-objective problem defined by Eq.~(\ref{EQOCPeq02}).

\section{Results and Discussion} \label{eq:results}

\subsection{Inverse Problem}

In order to apply the proposed methodology to solve the inverse problem described previously, the following steps are established:
\begin{itemize}
  \item Objective function: minimize the functional $\mathcal{F}$, given by Eq.~(\ref{EQInv01});
  \item Design space: $ 0.1 \le \beta \le 0.6 $, $ 0.04 \le \gamma \le 0.6 $ and $ 10^{-8} \le I_0 \le 0.5 $ (all defined after preliminary executions);
  \item DE parameters: population size (25), number of generations (100), perturbation rate (0.8), crossover rate (0.8) and strategy rand/1 (as presented in Section~\ref{sec:DE}). The evolutionary process is halted when a prescribed number of generations is reached (in this case, 100). Twenty independent runs of the algorithm were made, with different seeds for the generation of the initial population;
  \item To evaluate the SIR model during the optimization process, the Runge-Kutta-Fehelberg method was used;
  \item Initial conditions: $S(0)=1-I_0$, $I(0)=I_0$, and $R(0)=0$. In this case, $ I_0 $ is chosen as the first reported data in relation to the number of infected individuals in the time series;
  \item The data used in the formulation of the inverse problem refer to the population of China, from January 22 to April 2, 2020, taken from \cite{JHDC}.
\end{itemize}

\begin{figure}[!ht]
\centering
\includegraphics[scale=0.3]{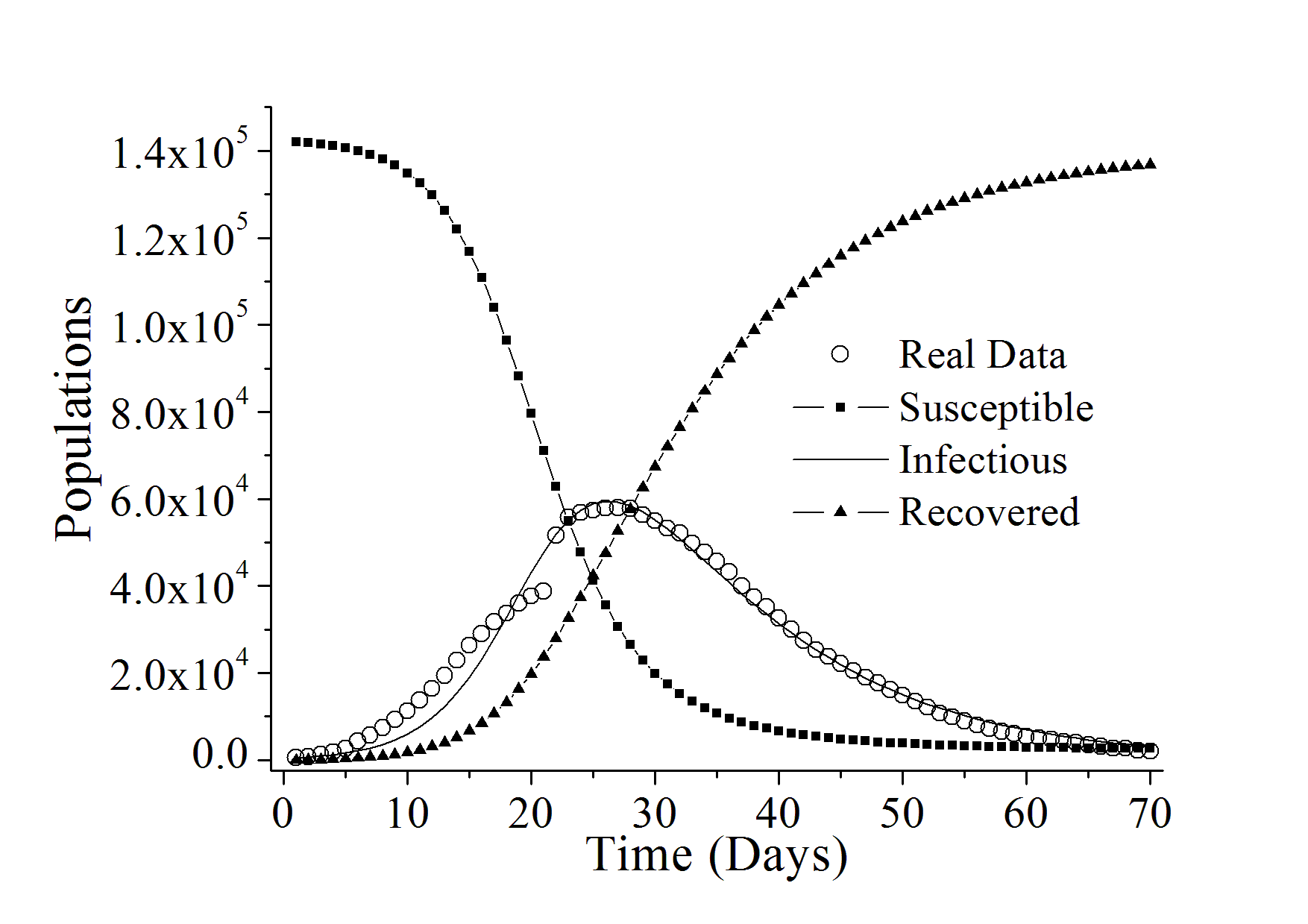}
\caption{Simulated and experimental profiles considering the estimated parameters.} \label{Fig2}
\end{figure}

Table~\ref{tabresinvprob} presents the results (best and standard deviation) obtained using DE. It is possible to observe that DE was able to obtain good estimates for the unknown parameters and, consequently, for the objective function, as can be verified, by visual inspection of Fig.~\ref{Fig2}. These results were obtained, as mentioned earlier, from 20 runs. Thus, the values of the standard deviation demonstrate that the algorithm converges, practically, to the same optimum in all executions (best). Physically, the probability of transmission by contact in the Chinese population is superior to 35 $\%$ ($\beta$ equal to 0.3566). In addition, $\gamma$ equal to 0.0858 implies a moderate \textit{per capita} recovery rate. One must consider that, since many cases may not be reported, for different reasons, as for example an asymptomatic infected person, the value of $ I_{0} $ may vary, as well as the behavior of the model over time.
\begin{table}[!ht]
\begin{center}
\caption{Results obtained for the proposed inverse problem considering DE.}\label{tabresinvprob}
{\setlength{\tabulinesep}{1.2mm}
\begin{tabu}{lcccc}
\hline
                    & $\beta \; \left( \mathrm{day^{-1}} \right)$ & $\gamma \; \left( \mathrm{day^{-1}} \right)$ & $I_0 \; \left( \textrm{Number of Individuals} \right) $ & $\mathcal{F}$\\ \hline
Best	            & 0.3566 & 0.0858 & 0.0038 & 0.1649 \\
Standard Deviation	& $ 1.2545 \times 10^{-5} $ & $ 1.6291 \times 10^{-5} $ & $ 1.43238 \times 10^{-6} $ & $ 1.2260 \times 10^{-7} $ \\\hline
\end{tabu}}
\end{center}
\end{table}

It is important to emphasize that when choosing $ I_0 $ as a design variable, the initial condition for the susceptible population ($ S_0 $) is automatically defined, that is, $ S_0 = 1 - I_0 $, since there is not, at the beginning of an epidemic, a considerable number of recovered individuals and, thus, $ R_0 $ = 0 is a reasonable choice. In this case, the available data refer to the number of infected individuals and these represent only the portion of individuals in the population that have actually been diagnosed. This is due, among other facts, to the lack of tests to diagnose the disease of all individuals who present symptoms. Thus, as the number of susceptible individuals at the beginning of the epidemic is dependent on the value of $ I_0 $, in this work it is considered that the total size of the population, typically defined as $ N = S + I + R $, is actually a portion of the total population, since the number of infected individuals available is also a fraction of those who have actually been diagnosed. In this case, the results presented below represent only the fraction of the infected population that was diagnosed and, consequently, the fraction of individuals susceptible to contracting the disease. Qualitatively, the results presented are proportional to the number of individuals in the population who were diagnosed with the disease.

In order to evaluate the sensitivity of the solutions obtained, in terms of the objective function, the best solution $ \left( \beta = 0.3566, \; \gamma = 0.0858, \; \textrm{and} \; I_0 = 0.0038 \right) $ was analyzed considering a perturbation rate given by $\delta$. For this purpose, the range $ \left[ \left( 1 - \delta \right) \theta_{k}, \; \left( 1 + \delta \right) \theta_{k} \right] $ was adopted, for $ k \subset \lbrace 1, \; 2, \; 3 \rbrace $, where $ \boldsymbol{\theta} = \left( \beta, \; \gamma, \; I_{0} \right) $. Thus, in each analysis, one design variable is perturbed and the value of $\mathcal{F}$ in relation to this noise is computed.

\begin{figure}[!ht]
\centerline{ \subfigure[$\gamma=0.0858$ and $I_0$=0.0038.]
{\includegraphics[scale=0.22]{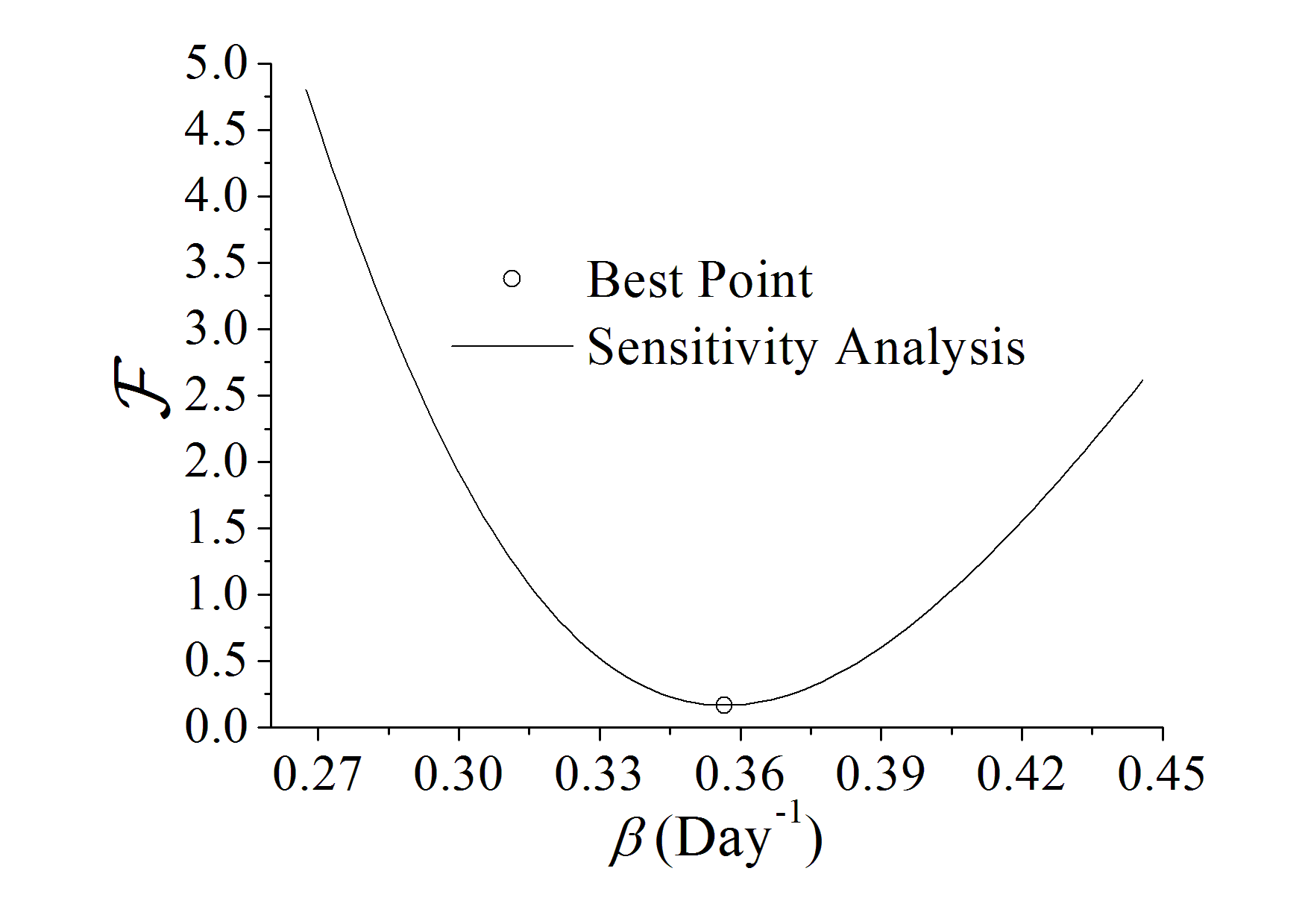}}
\hspace{-1.2em}
\hfill \subfigure[$\beta=0.3566$ and $I_0$=0.0038.]
{\includegraphics[scale=0.22]{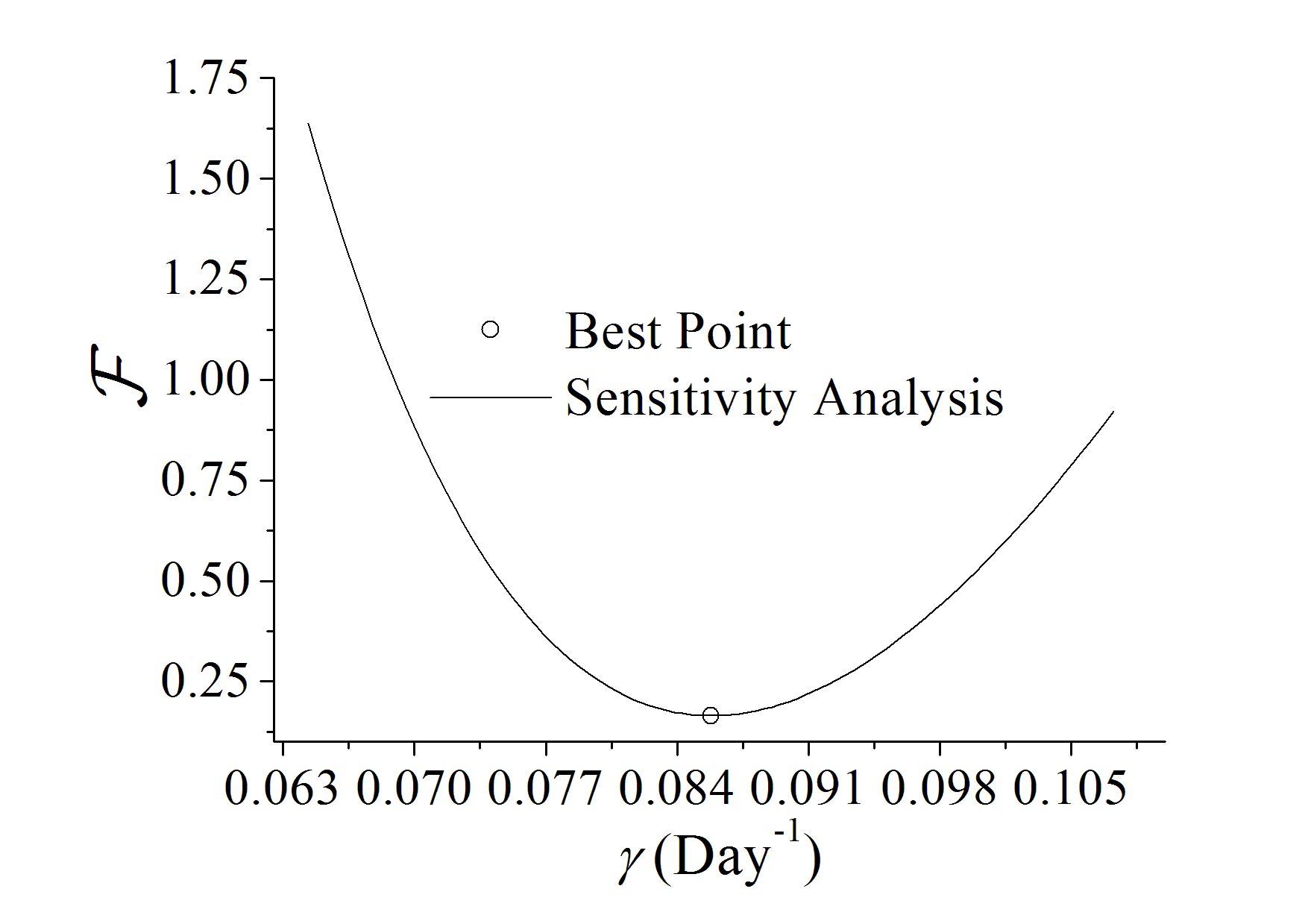}}
\hspace{-1.2em}
\hfill \subfigure[$\beta=0.3566$ and $\gamma=0.0858$.]
{\includegraphics[scale=0.22]{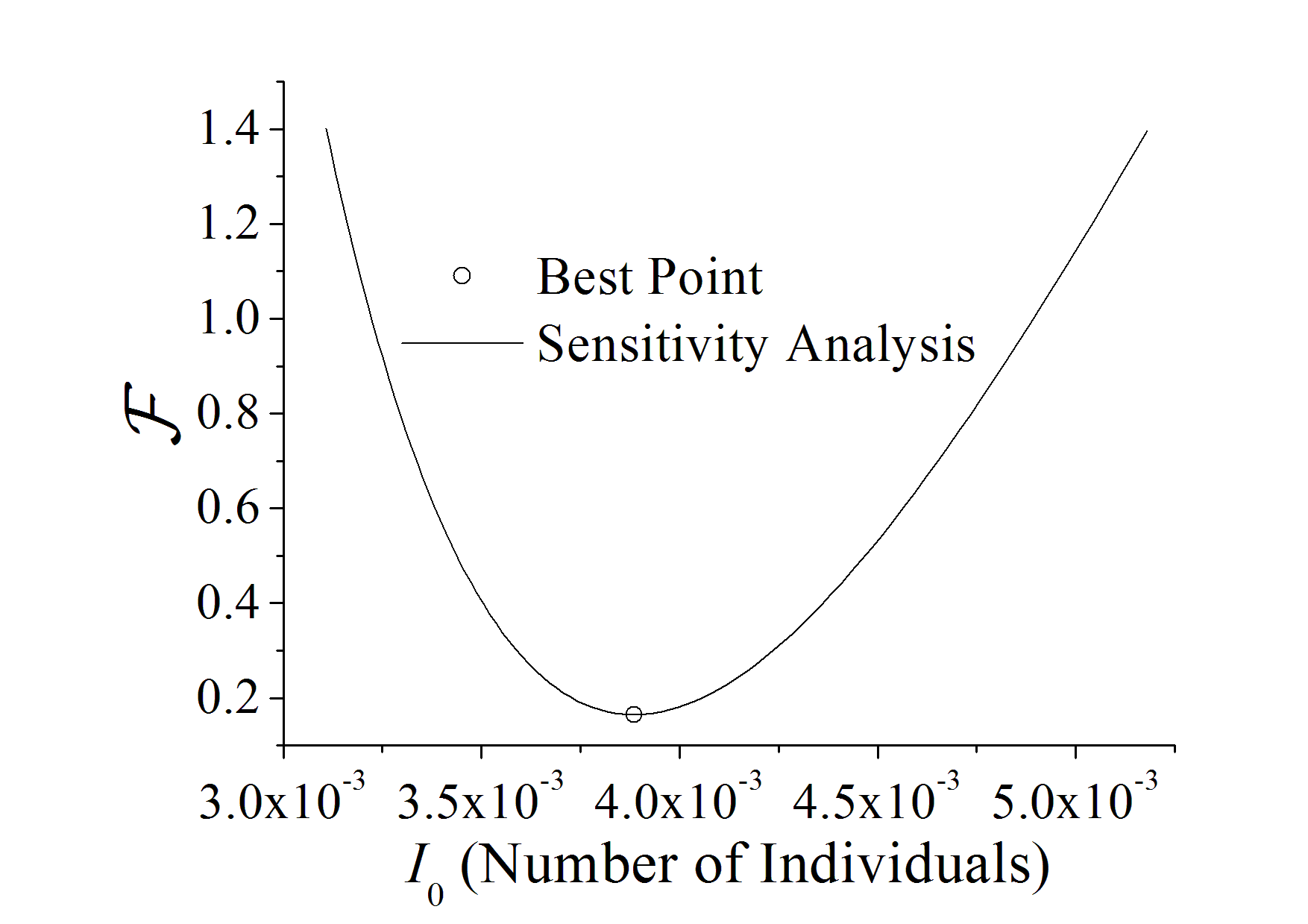}}}
\caption{Sensitivity analysis of estimated parameters.}
\label{Fig3}
\end{figure}

Figure~\ref{Fig3} presents the sensitivity analysis for each estimated parameter, in terms of the objective function, considering $\delta$ equal to 0.25 and 100 equally spaced points in the interval of interest. In these figures, it is possible to observe that the variation of each parameter, as expected, in a worst value for the $\mathcal{F}$. In addition, that the design variable more sensible to $\delta$ parameter is the $\beta$ parameter, since a wider range of values for the $\mathcal{F}$ were obtained.

\subsection{Mono-objective Optimal Control Problem}

We consider two distinct analysis in this section, in order to evaluate the proposed methodology considered to solve the mono-objective optimization problem: \textit{i}) solution of the proposed mono-objective optimal control problem and; \textit{ii}) evaluation on the influence of the maximum amount of vaccine, by defining an inequality constraint. For this purpose, the following steps are established:
\begin{itemize}
  \item Objective function: minimize the functional $\Omega_1$, given by Eq.~(\ref{EQOCPeq01});
  \item The previously calculated parameters ($\beta$, $\gamma$ and $I_0$) are employed in the simulation of the SIR model;
  \item Design space: $ 0 \le t_i \le t_f$, for $i = 1, \; \hdots,  \; t_{N_{\textrm{elem}}-1}$, and $N_{\textrm{elem}} = 10 $. It is important to mention that this value was chosen after preliminary runs, i.e., increasing this value do not produce better results in terms of the objective function;
  \item DE parameters: population size (25), number of generations (100), perturbation rate (0.8), crossover rate (0.8) and strategy rand/1 (as presented in Section~\ref{sec:DE}). The evolutionary process is halted when a prescribed number of generations is reached (in this case, 100). 20 independent runs of the algorithm were made, with different seeds for the generation of the initial population;
  \item To evaluate the SIR model during the optimization process, the Runge-Kutta-Fehelberg method was used;
  \item Initial conditions: $S(0)=1-I_0$, $I(0)=I_0$, and $R(0)=0$. As in the previous case, $ I_0 $ is chosen as the first reported data in relation to the number of infected individuals in the time series;
\end{itemize}

\begin{figure}[!ht]
\centerline{ \subfigure[Objective Function.]
{\label{Fig4newa}\includegraphics[scale=0.22]{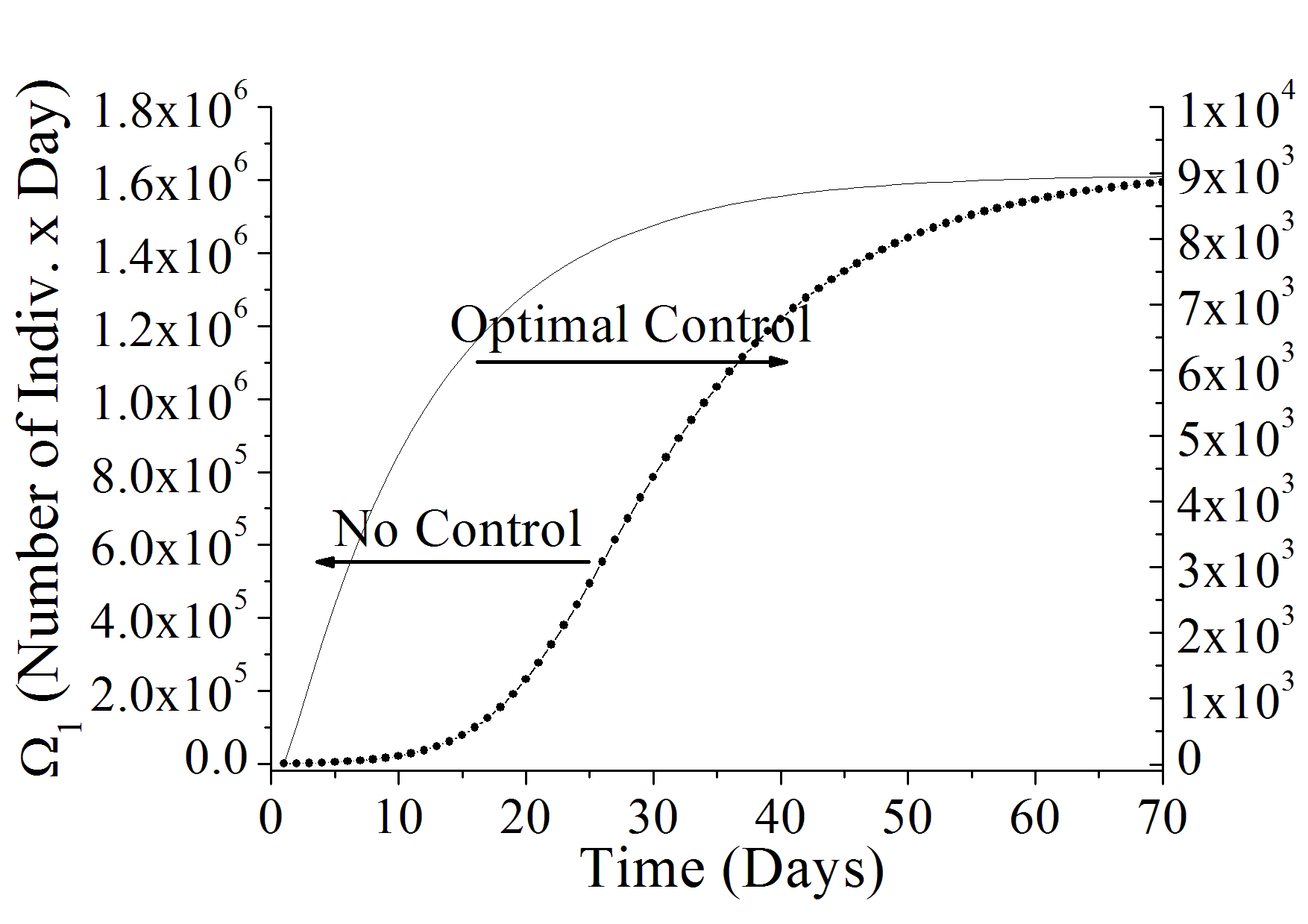}}
\hspace{-0.5cm}
\hfil \subfigure[Susceptible.]
{\label{Fig4newb}\includegraphics[scale=0.22]{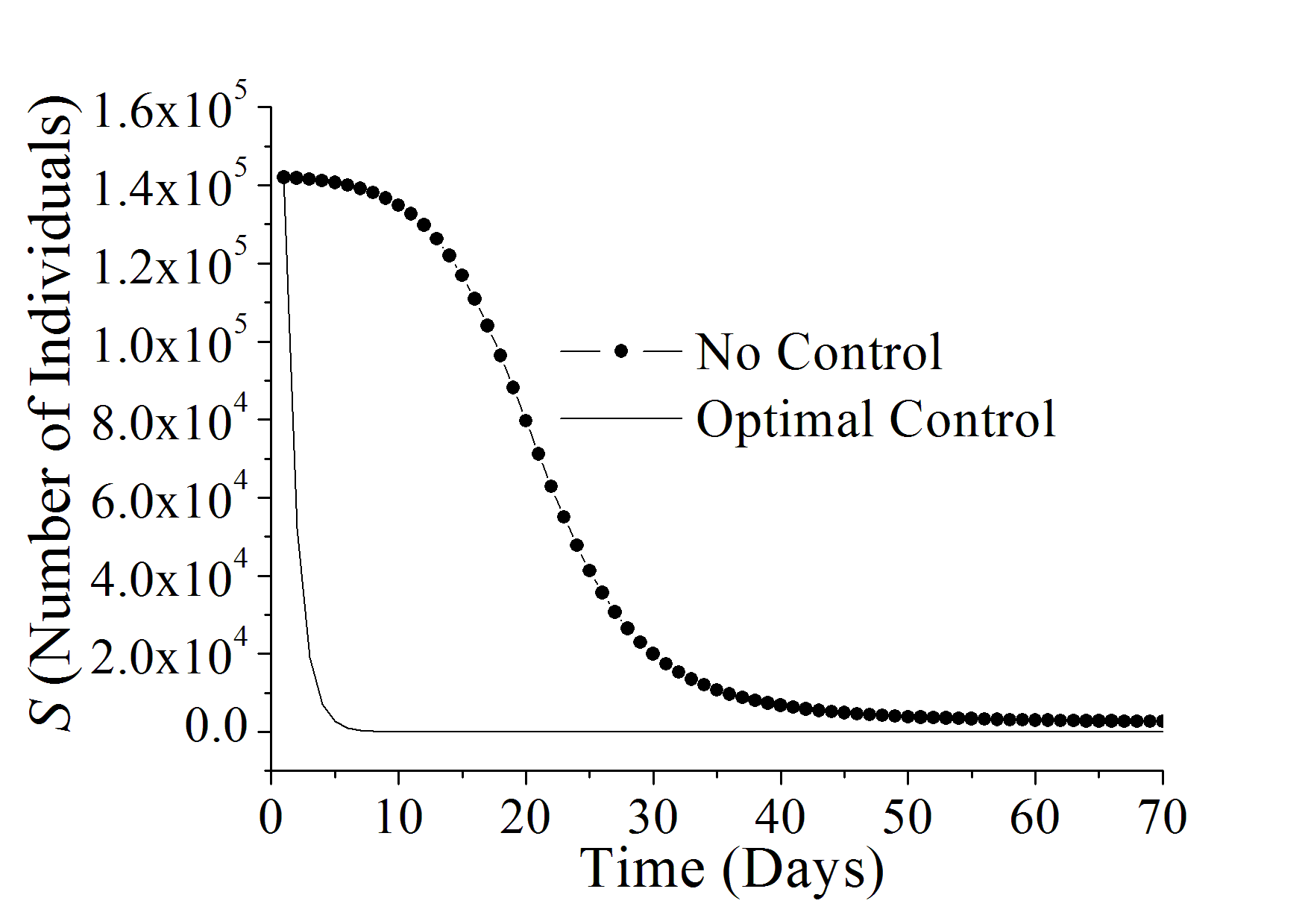}
}}
\vspace{-0.25cm}
\centerline{ \subfigure[Infectious.]
{\label{Fig4newc}\includegraphics[scale=0.22]{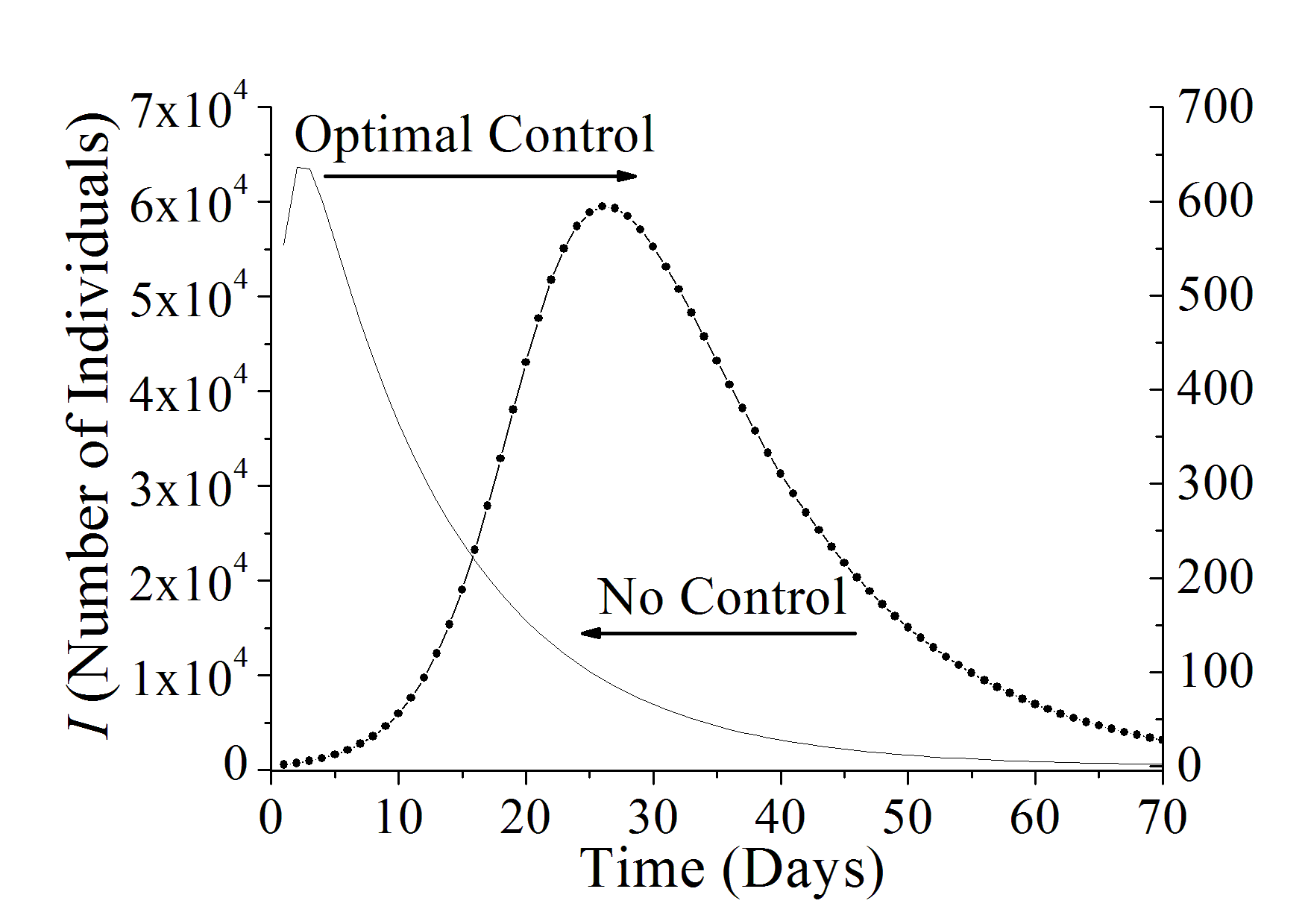}}
\hspace{-0.5cm}
\hfil \subfigure[Recovered.]
{\label{Fig4newd}\includegraphics[scale=0.22]{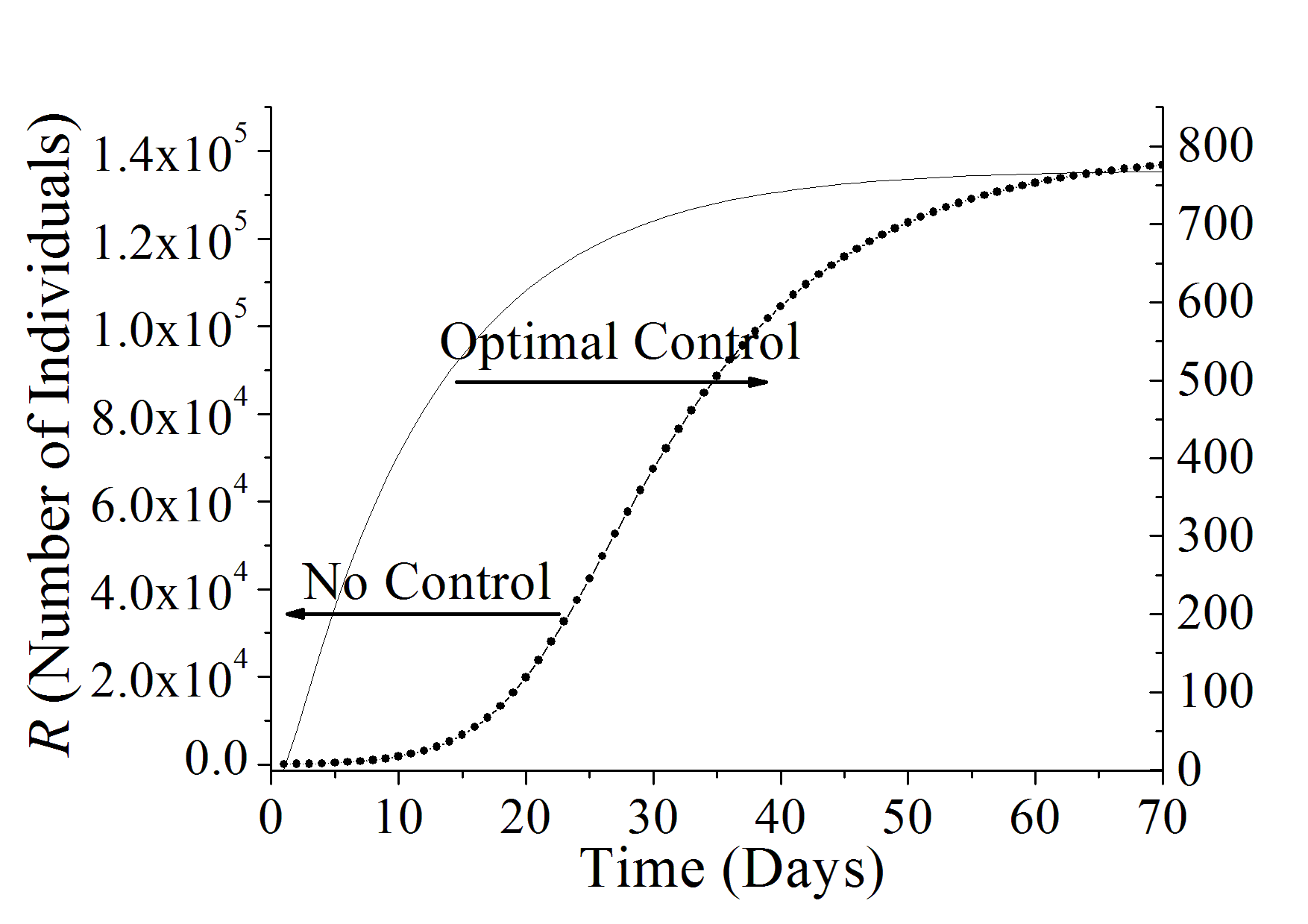}
}}
\vspace{-0.25cm}
\centerline{ \subfigure[Control Variable.]
{\label{Fig4newe}\includegraphics[scale=0.22]{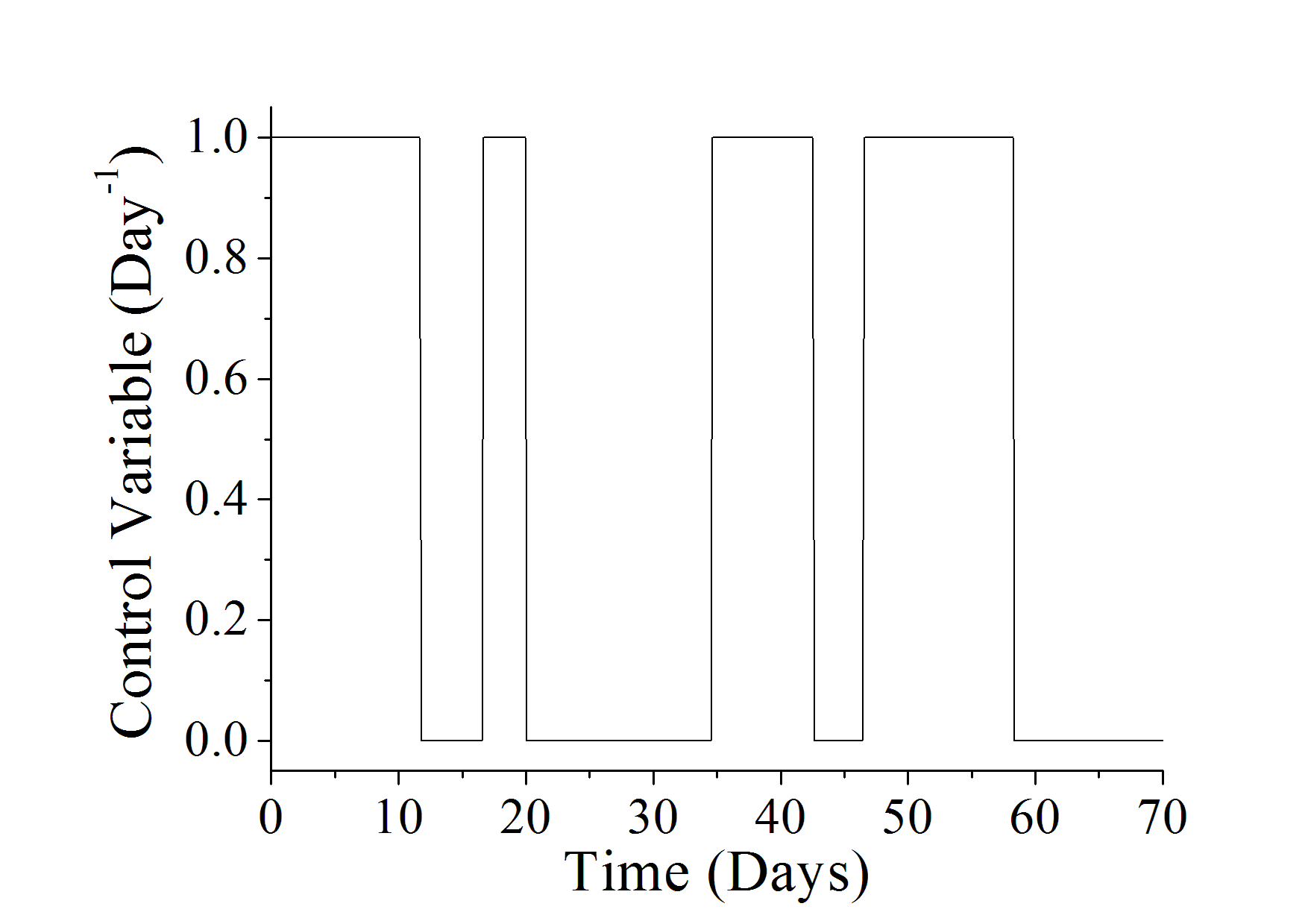}}
\hspace{-0.5cm}
\hfil \subfigure[Number of Vaccinated Individuals.]
{\label{Fig4newf}\includegraphics[scale=0.22]{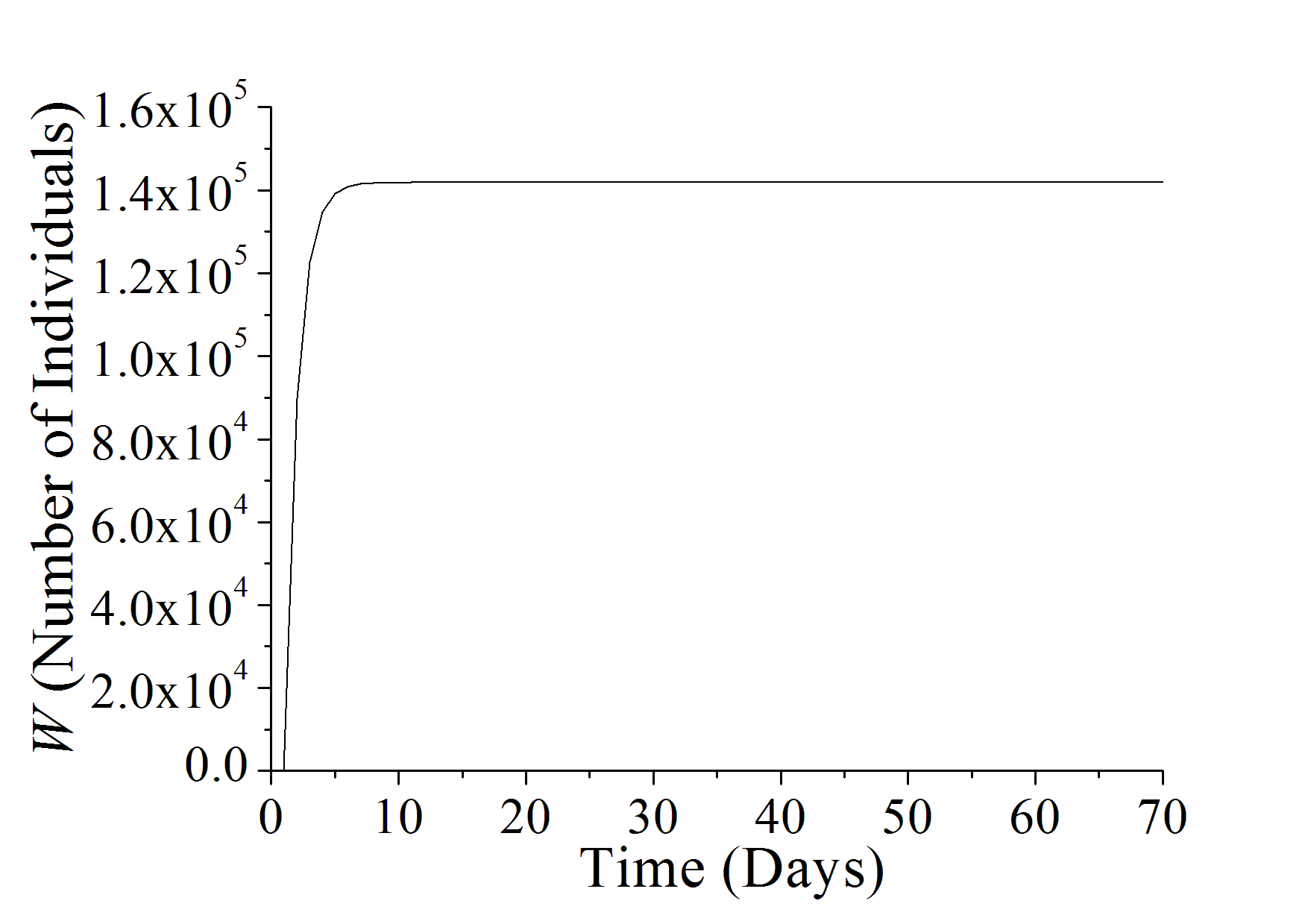}
}}
\caption{Objective function, susceptible-infectious-removed populations profiles, control variable strategy and number of vaccinated individuals' profiles.} \label{Fig4new}
\end{figure}

Table~\ref{tabocp01} presents the best solution obtained by using DE and considering ten control elements, in terms of the number of individuals. The objective function obtained (about 8945.4278 individuals) is less than the case in which no control is considered (about 1594607.2234 individuals), i.e., the number of infected individuals is lower when a control strategy is considered (see Figs.~\ref{Fig4newa} and \ref{Fig4newc}). If the number of infected individuals is reduced, due to control action, the number of susceptible individuals rapidly decreases until its minimum value ($ 1.4382 \times 10^{-3}$) and, consequently, the number of recovered individuals rapidly increase until its maximum value (767.5187 individuals), as observed in Figs.~\ref{Fig4newb} and \ref{Fig4newd}, respectively. In terms of the action regarding the control variable, the effectiveness is readily verified in the beginning of the vaccine administration. Further the administration is conducted in specific intervals of time, which preserves the health of the population, as observed in Fig.~\ref{Fig4newe}. The evolution of the number of vaccinated individuals is presented in Fig.~\ref{Fig4newf}. In this case, due to control action, the vaccinated population increase rapidly until the value is saturated (141835.1405). In summary, all obtained profiles are coherent from the physical point of view. Finally, it is important to mention that the standard deviation for each result is, approximately, equal to $10^{-3}$, which demonstrates the robustness of DE to solve the proposed mono-objective optimal control problem.
\begin{table}[!ht]
\begin{center}
\caption{Results obtained for the proposed mono-objective optimization problem ($t_f=70$ days).}\label{tabocp01}
{\setlength{\tabulinesep}{1.2mm}
\begin{tabu}{ccccc}
\hline
$\Omega_1$ (Number of Individuals $\times$ Days) & $S(t_f)$   & $I(t_f)$ & $R(t_f)$ & $W(t_f)$\\\hline
8945.4278    & 1.4382E-03    &  2.1201 &  767.5187 & 141835.1405 \\\hline
\end{tabu}}
\end{center}
\end{table}

\begin{figure}[!ht]
\centerline{ \subfigure[Objective Function.]
{\includegraphics[scale=0.22]{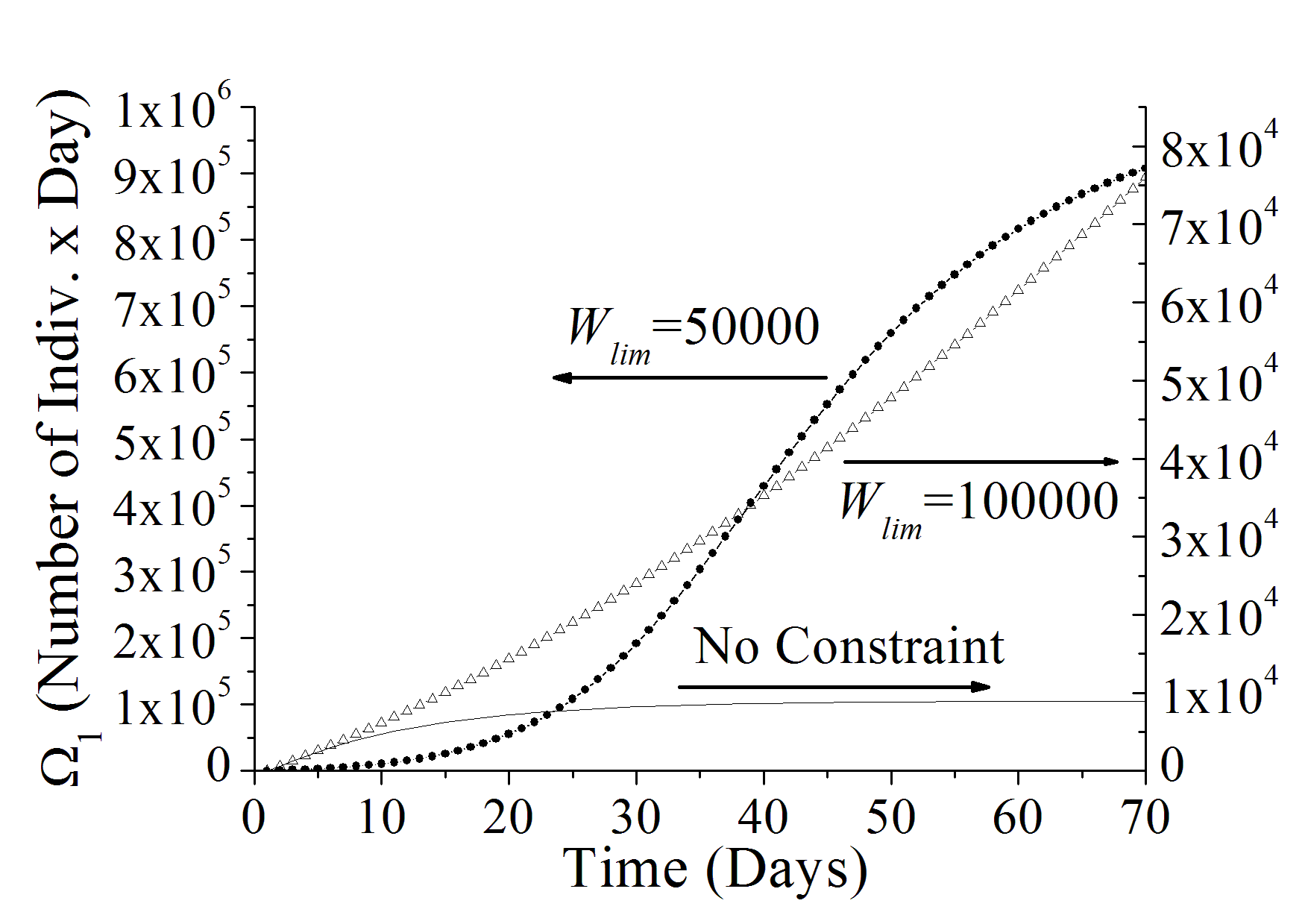}}
\hspace{-0.5cm}
\hfil \subfigure[Susceptible.]
{\includegraphics[scale=0.22]{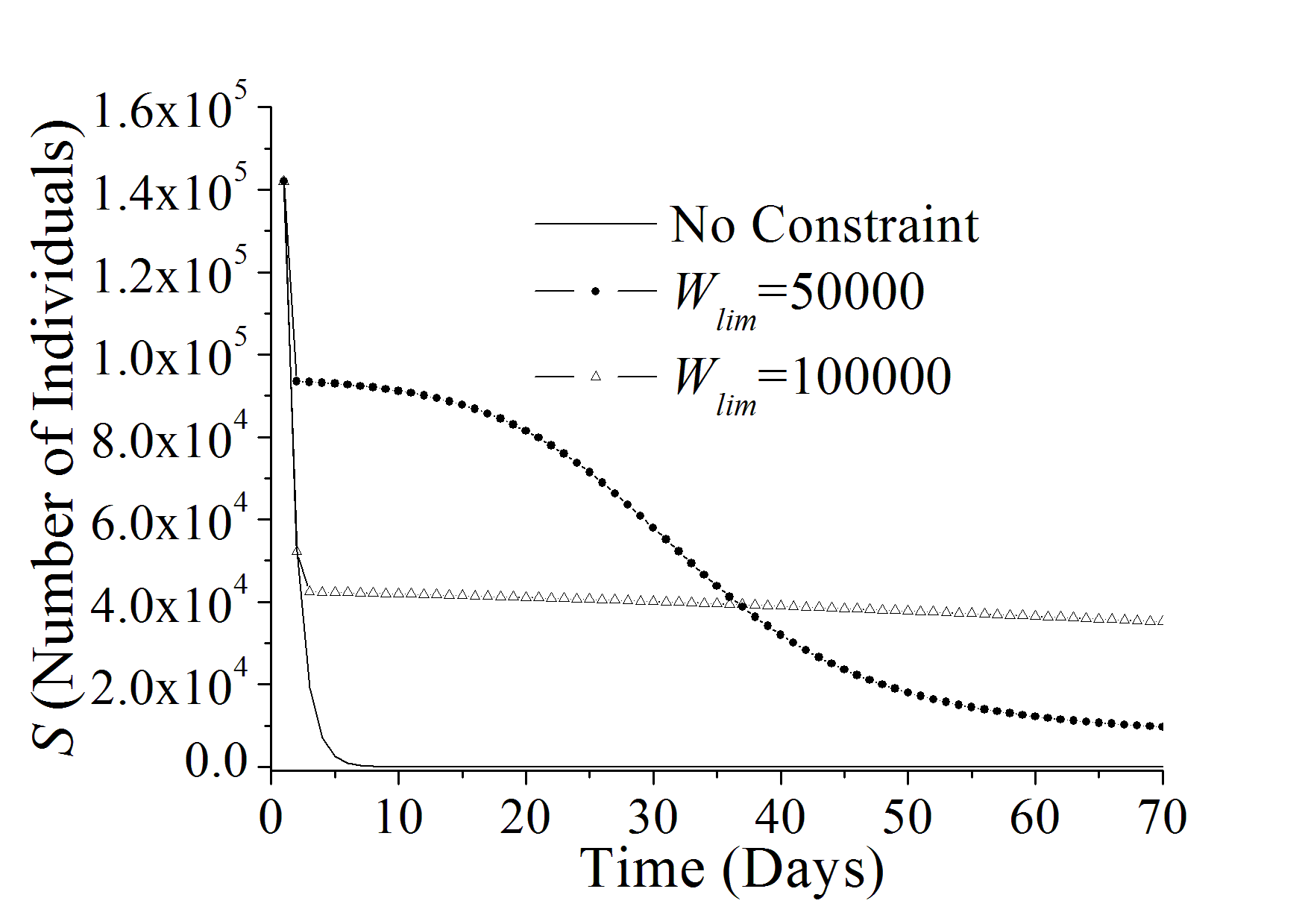}
}}
\vspace{-0.25cm}
\centerline{ \subfigure[Infectious.]
{\includegraphics[scale=0.22]{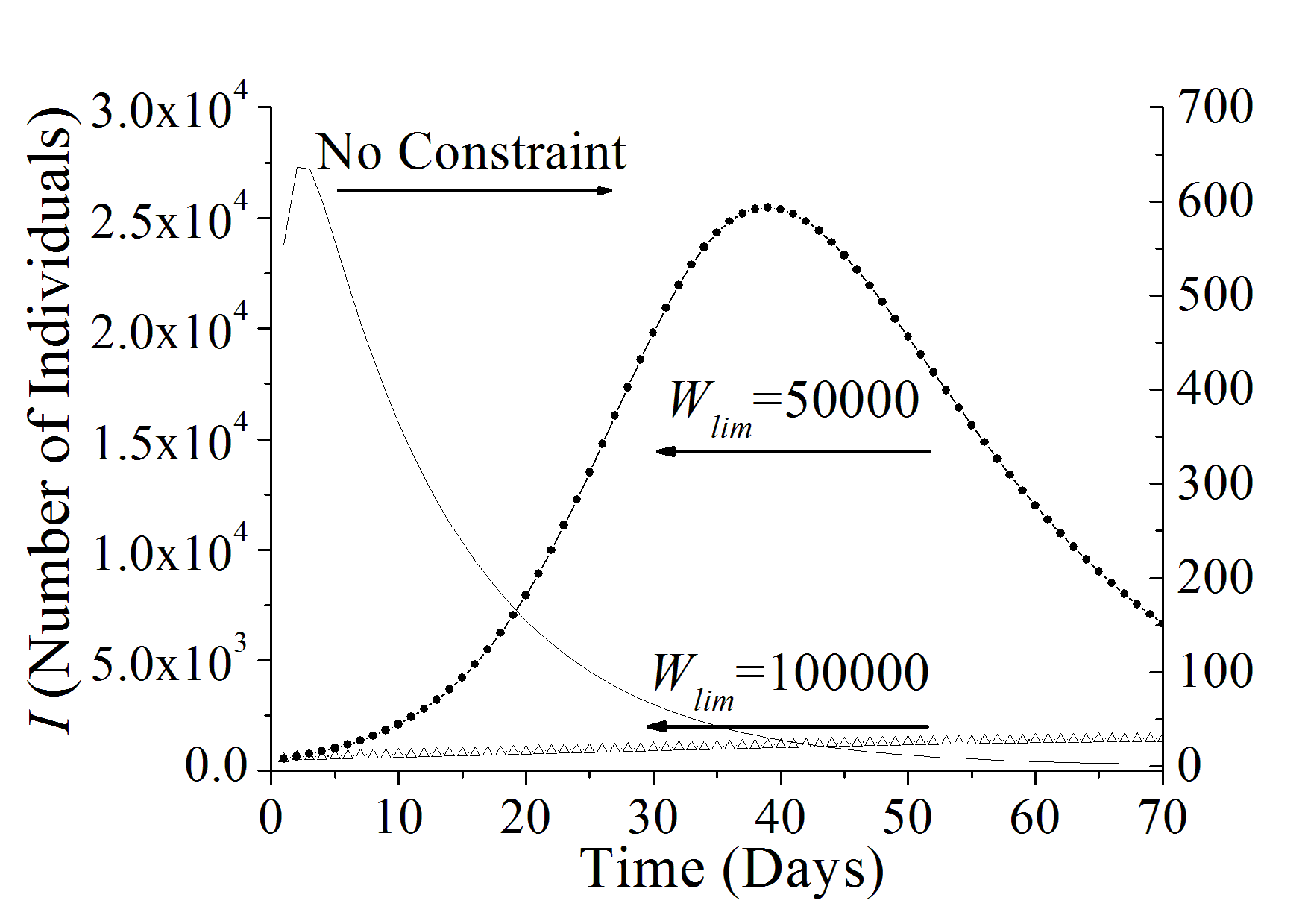}}
\hspace{-0.5cm}
\hfil \subfigure[Recovered.]
{\includegraphics[scale=0.22]{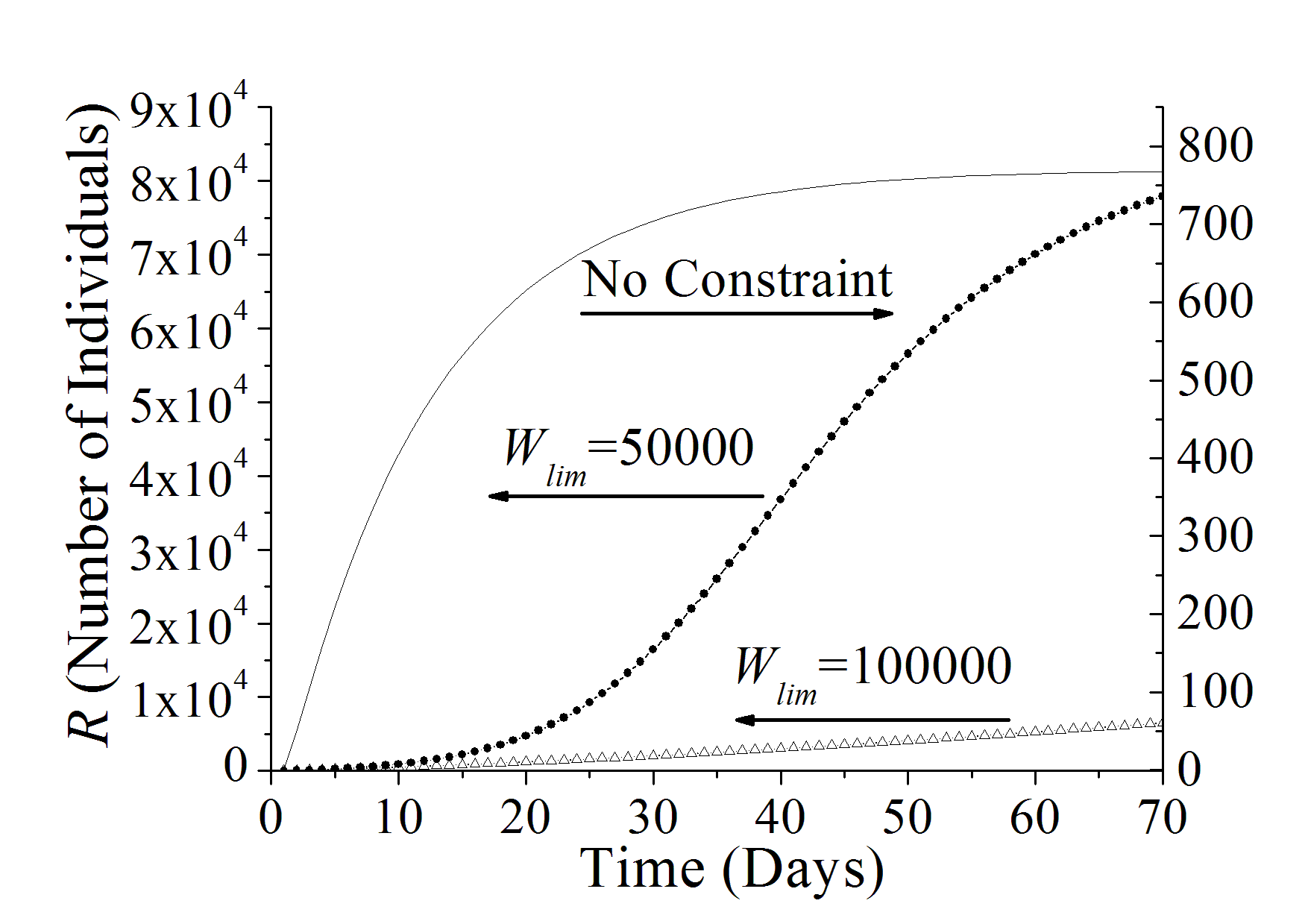}
}}
\vspace{-0.25cm}
\centerline{ \subfigure[Control Variable.]
{\includegraphics[scale=0.22]{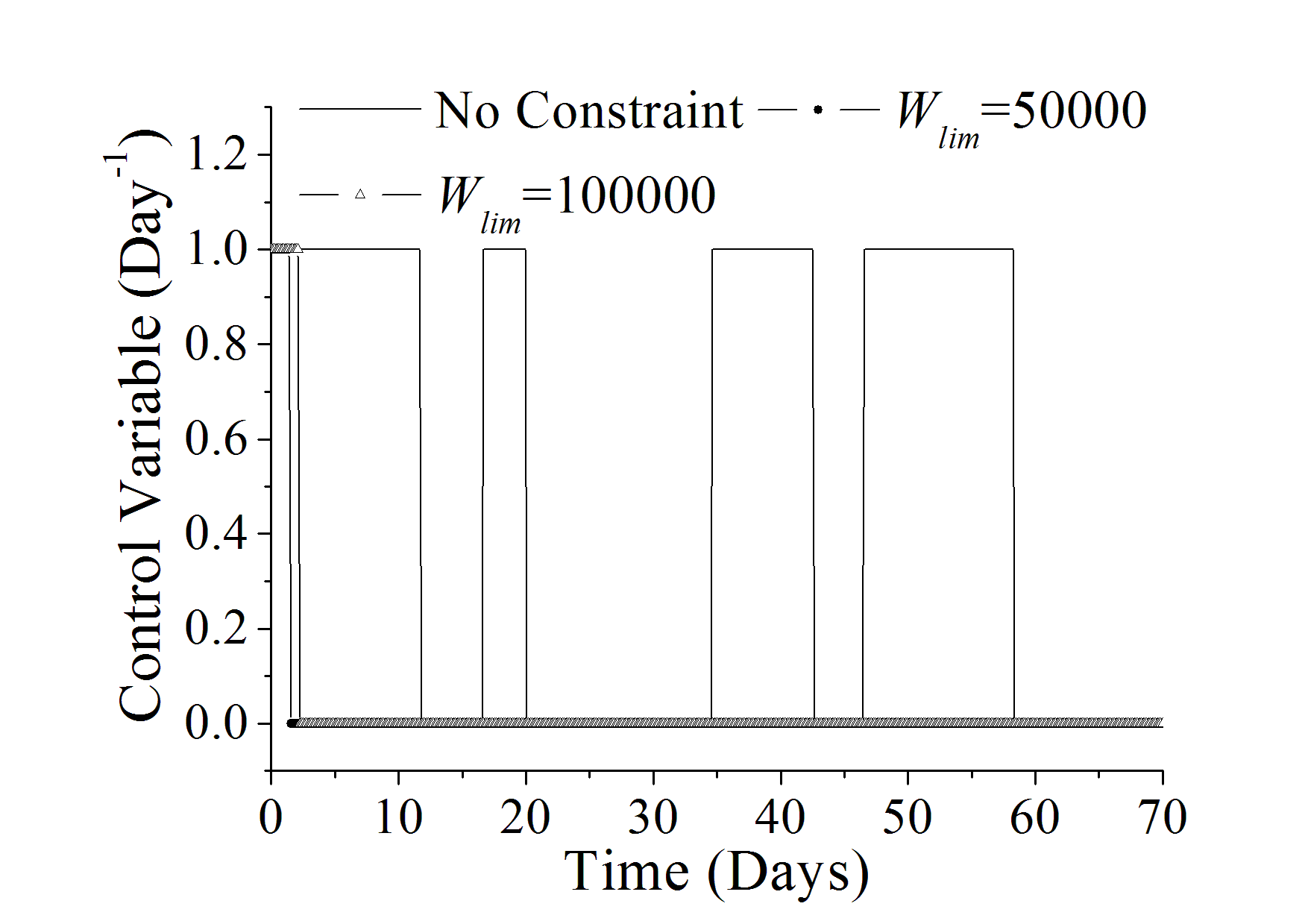}}
\hspace{-0.5cm}
\hfil \subfigure[Number of Vaccines.]
{\includegraphics[scale=0.22]{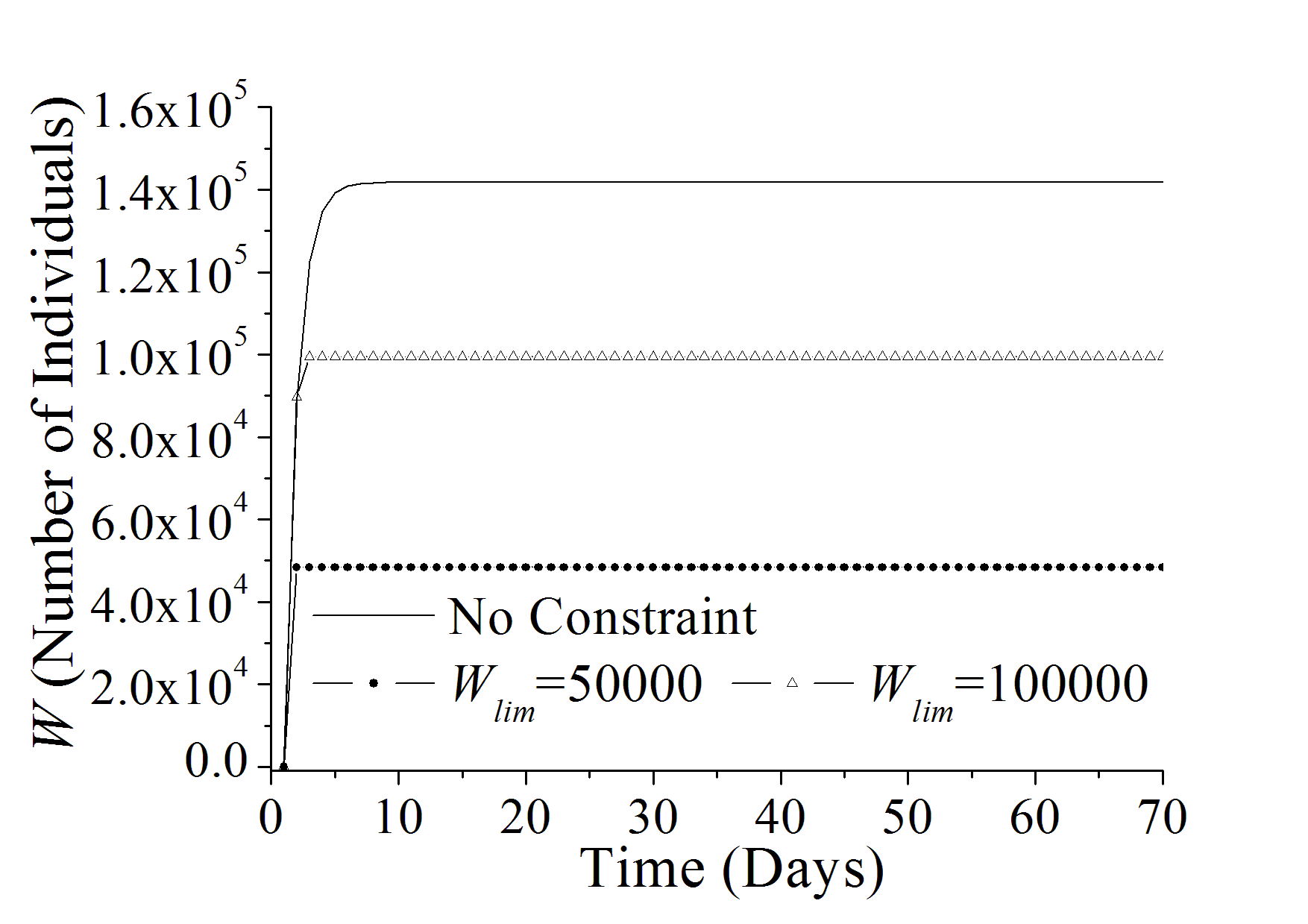}
}}
\caption{Influence of the maximum amount of vaccine in the objective function, susceptible-infectious-removed populations profiles, control variable strategy and number of vaccinated individuals' profiles.} \label{Fig6new}
\end{figure}

In this model, the evaluation of the number of vaccinated individuals is associated with an inequality constraint. This relation bounds the quantity of individuals that can be vaccinated due to the limitation related to the production of vaccines. For this purpose, two control elements are incorporated to the model: if $W(t_1) \le W_{\mathrm{lim}}$, then $u=1$. Otherwise, $u=0$ ($t_1$ is the instant of time that $W(t_1)=W_{\mathrm{lim}}$, and $W_{\mathrm{lim}}$ is the upper bound for the number of vaccinated individuals). Table~\ref{tabocp02} presents the results obtained considering different quantities for the parameter $W_{\mathrm{lim}}$. As expected, the insertion of this constraint implies in limiting the maximum number of vaccinated individuals and, consequently, a lower number of individuals are vaccinated. The increase of the parameter $W_{\mathrm{lim}}$ implies in the reduction of the objective function value, in number of infected and recovered individuals and, consequently, an increase in the number of susceptible individuals. These analysis can be observed in Fig.~\ref{Fig6new}.
\begin{table}[!ht]
\begin{center}
\caption{Results obtained for the proposed mono-objective optimization problem considering different quantities for the parameter $W_{\mathrm{lim}}$ ($t_f=70$ days).}\label{tabocp02}
{\setlength{\tabulinesep}{1.2mm}
\begin{tabu}{cccccc}
\hline
\ $W_{\mathrm{lim}}$ & \ $t_1$ (Days) & \ $\Omega_1$ (Number of Individuals $\times$ Days) & \ $S(t_f)$  & \ $I(t_f)$ & \ $R(t_f)$ \\ \hline
\  50000  & \  1.4389  & \  907790.2114   &  \ 9674.7066   & \   6659.1695       & \  77888.4004          \\
\  100000  & \ 2.1985   & \  76039.6424   &  \ 35197.4174  & \   1472.4917       & \  6524.2013          \\\hline
\end{tabu}}
\end{center}
\end{table}

\newpage
\subsection{Multi-objective Optimal Control Problem}

As presented previously, a multi-objective optimal control problem was proposed in order to minimize the number of infected individuals ($\Omega_1$) and to minimize the quantity of vaccine administered ($\Omega_2$). To evaluate the proposed methodology considered to solve this multi-objective optimization problem, the following steps are established:
\begin{itemize}
  \item Objective functions: minimize both $\Omega_1$ and $\Omega_2$ together, which are defined by Eq.~(\ref{EQOCPeq02});
  \item The previously calculated parameters ($\beta$, $\gamma$ and $I_0$) are employed in the simulation of the SIR model;
  \item Design space: $ 0 \le t_i \le t_f$, for $i = 1, \; \hdots,  \; t_{N_{\textrm{elem}}-1}$, and $N_{\textrm{elem}} = 10 $.
  \item MODE parameters: population size (50), number of generations (100), perturbation rate (0.8), crossover rate (0.8), number of pseudo-curves (10), reduction rate (0.9), and strategy rand/1 (as presented in Section~\ref{sec:DE}). The stopping criterion adopted is the same as in the previous cases.
  \item To evaluate the SIR model during the optimization process, the Runge-Kutta-Fehelberg method was used;
  \item Initial conditions: $S(0)=1-I_0$, $I(0)=I_0$, $R(0)=0$, and $W(0)=0$.
\end{itemize}

\begin{figure}[!ht]
\centerline{ \subfigure[Pareto curve.]
{\label{Fig7newa}\includegraphics[scale=0.22]{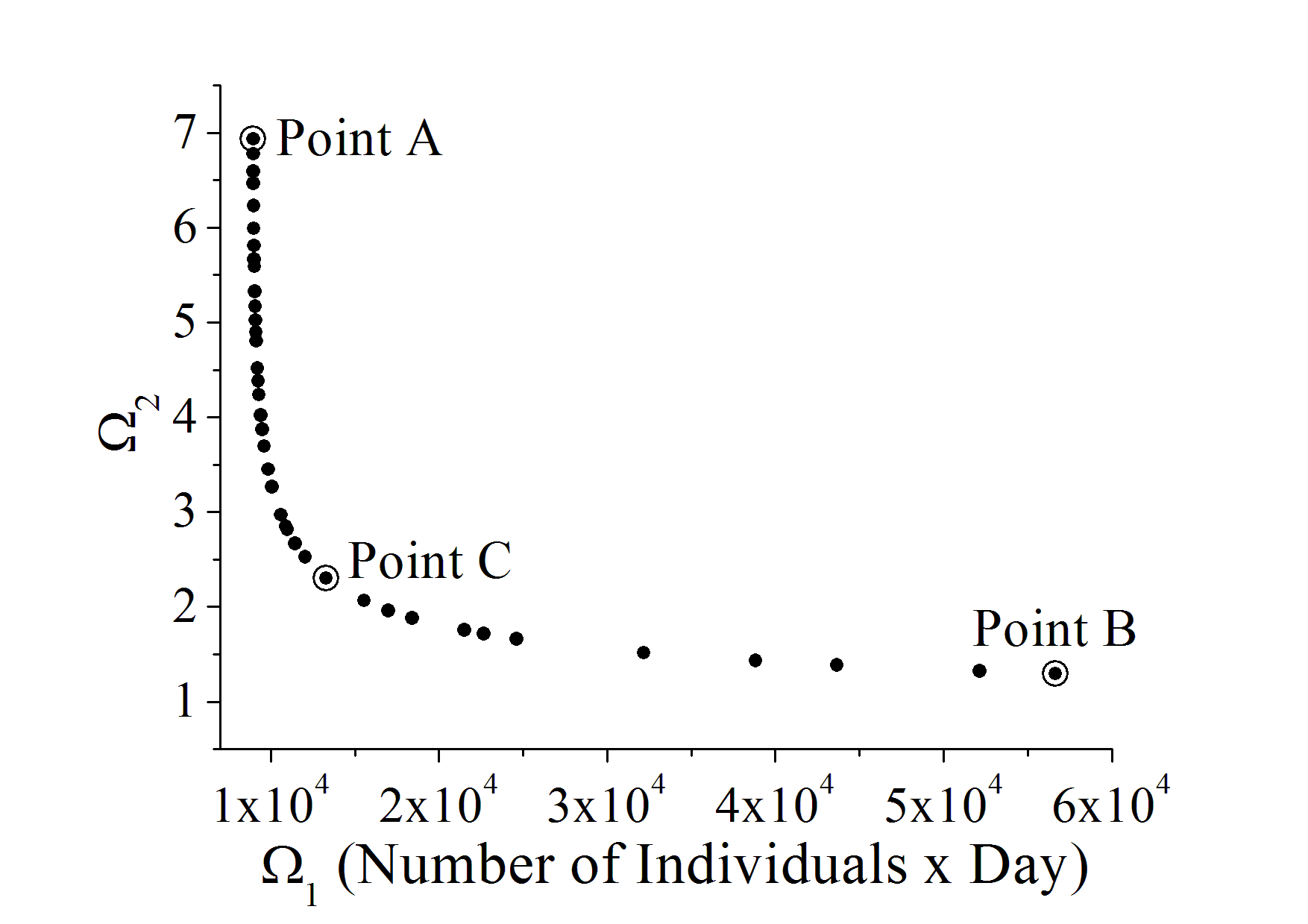}}
\hspace{-0.5cm}
\hfil \subfigure[Susceptible.]
{\label{Fig7newb}\includegraphics[scale=0.22]{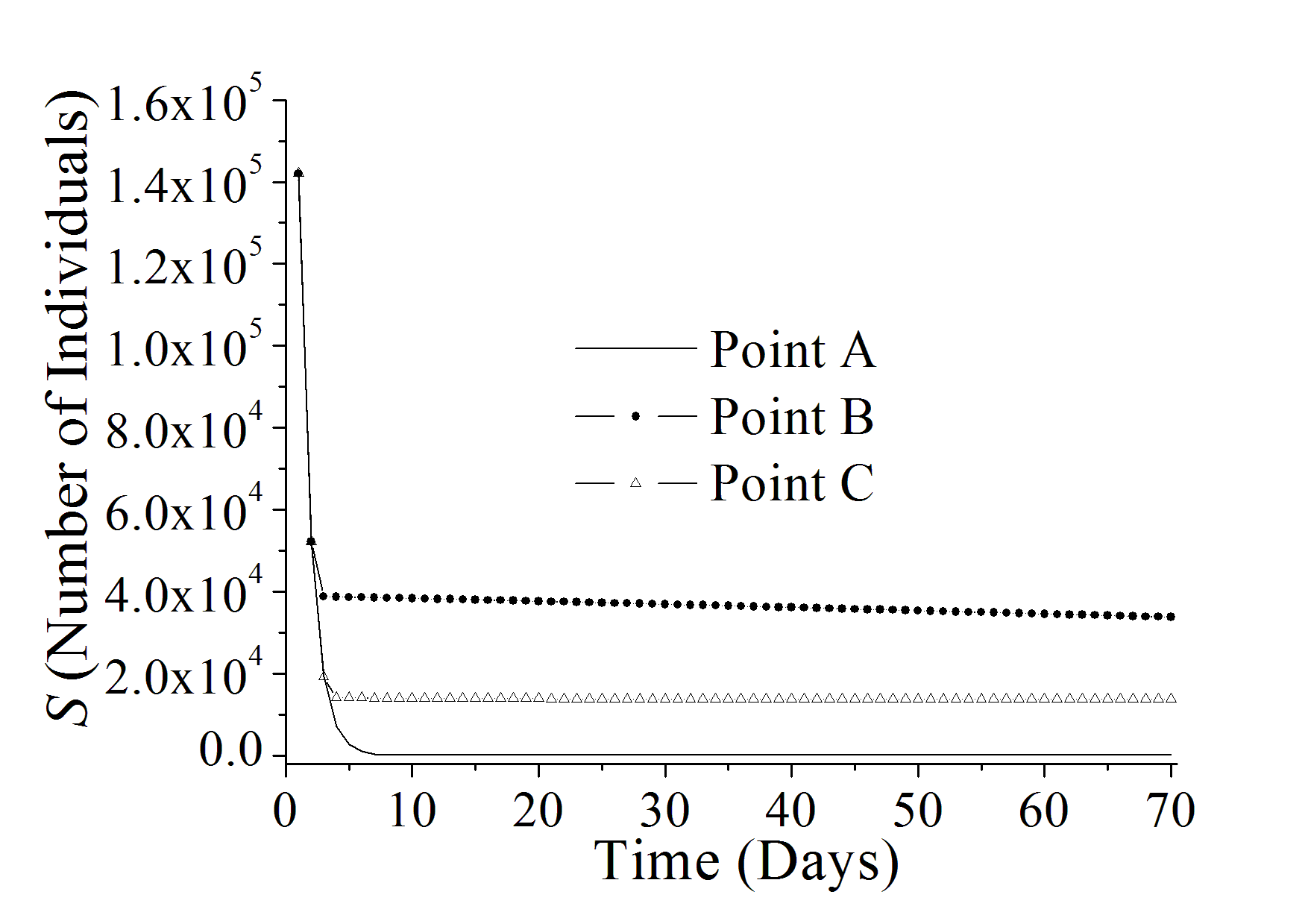}
}}
\vspace{-0.25cm}
\centerline{ \subfigure[Infectious.]
{\label{Fig7newc}\includegraphics[scale=0.22]{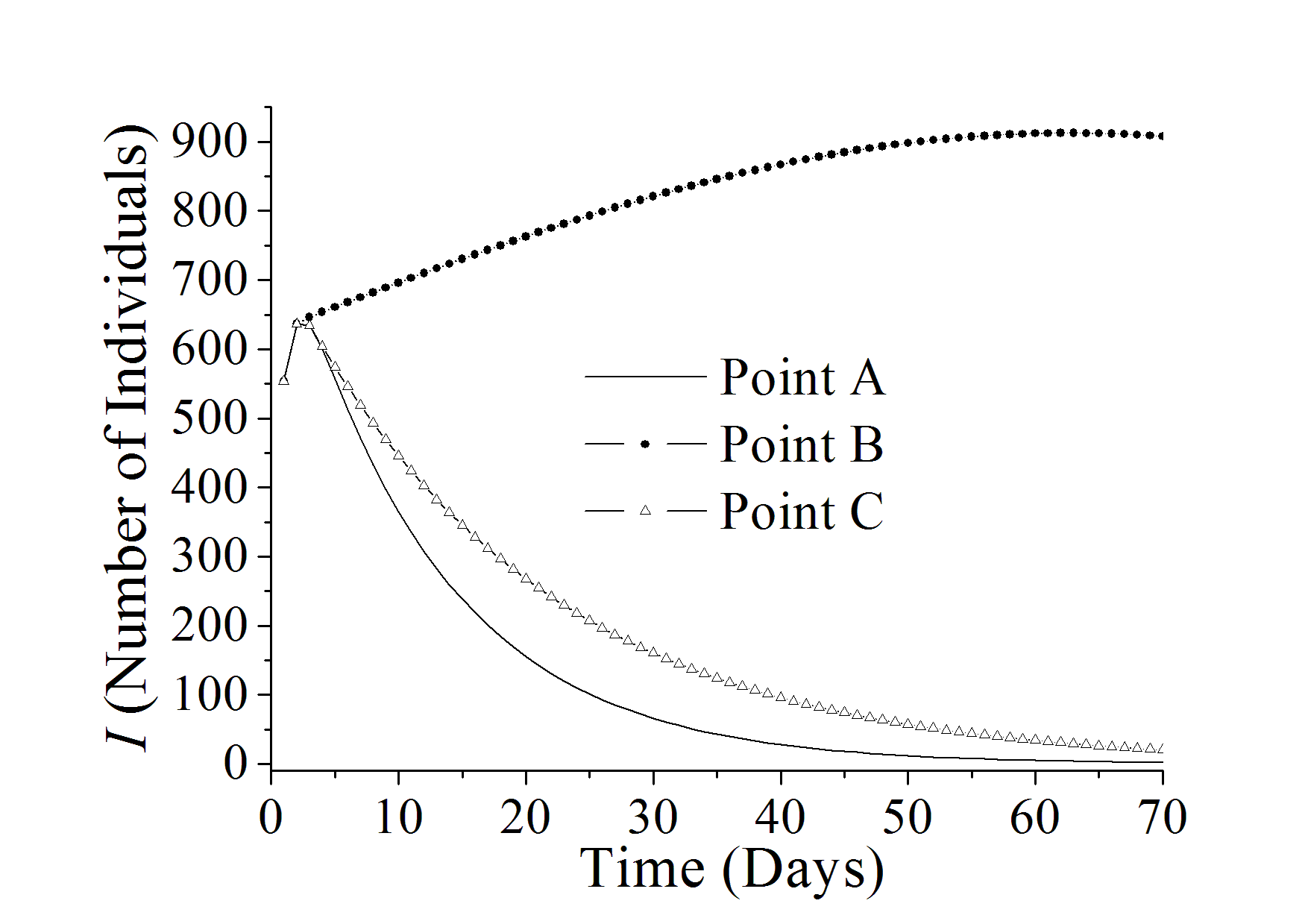}}
\hspace{-0.5cm}
\hfil \subfigure[Recovered.]
{\label{Fig7newd}\includegraphics[scale=0.22]{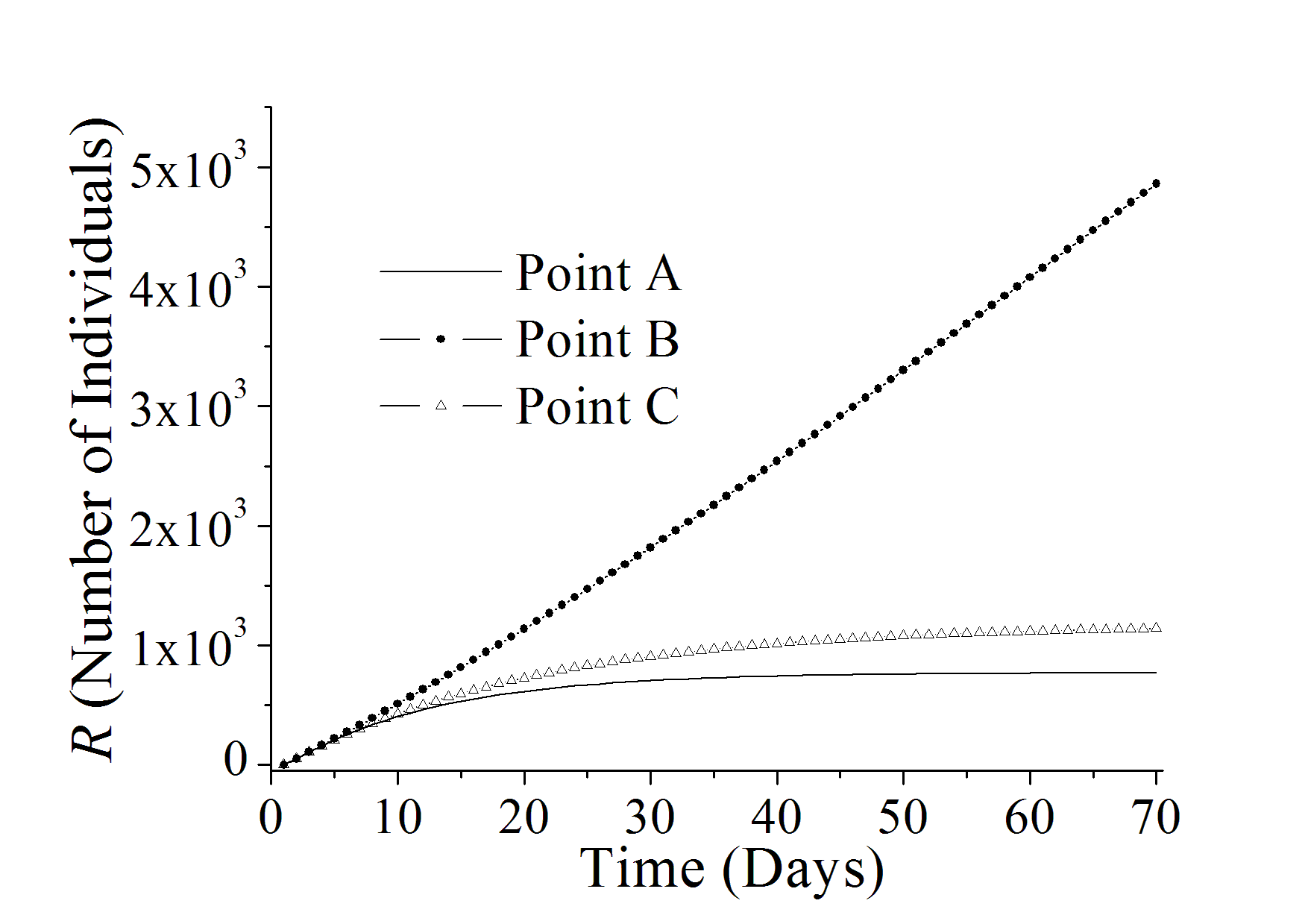}
}}
\vspace{-0.25cm}
\centerline{ \subfigure[Control Variable.]
{\label{Fig7newe}\includegraphics[scale=0.22]{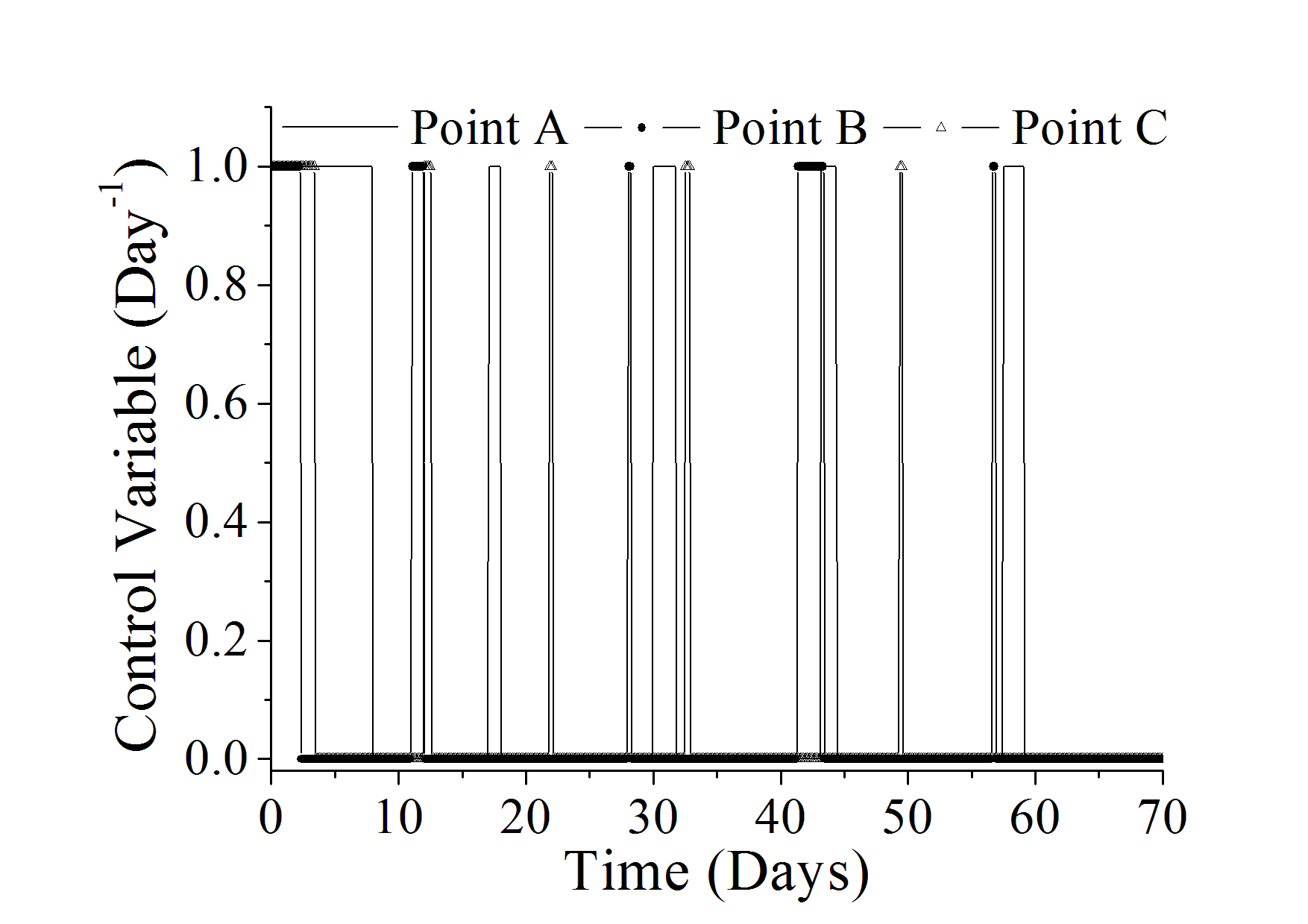}}
\hspace{-0.5cm}
\hfil \subfigure[Number of Vaccines.]
{\label{Fig7newf}\includegraphics[scale=0.22]{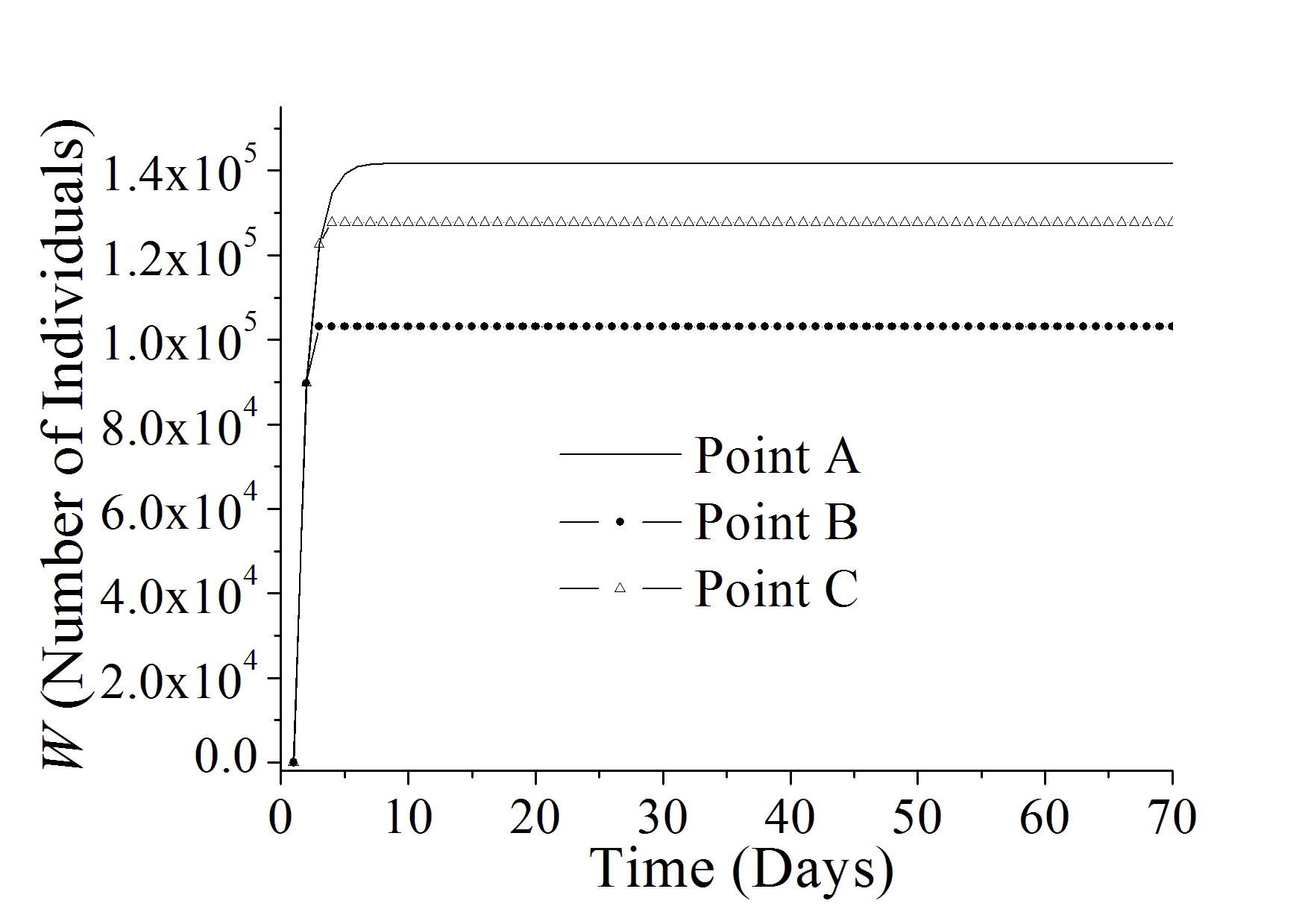}
}}
\caption{Pareto curve, susceptible-infectious-removed populations profiles, control variable strategy and number of vaccinated individuals' profiles.} \label{Fig7new}
\end{figure}

Figure~\ref{Fig7newa} presents the Pareto curve and three points (A, B and C) belonging to this curve, as shown in Table~\ref{tabpareto}. It must be stressed that the Pareto curve presents the non-dominated solutions, as described in Section~\ref{sec:MODE}. The point A represents the best solution in terms of the minimization of the number of infected individuals, with $\Omega_1=8963.7775$, that is, the number of infected individuals at $t_f$ assume its lowest value, which is equal to $I(t_f)=769.0921$, but considering a larger amount of vaccine administered ($\Omega_2=6.9358$). On the other hand, the point B represents the best solution in terms of the quantity of vaccine administered, with $\Omega_2=1.2940$, i.e, the minimization of such value when $t=t_f$. However, for this point, the number of infected individuals is high ($\Omega_1=56644.0350$). The point C is a compromise solution, which is a good solution in terms of both objectives simultaneously, with intermediary values for both objectives, $\Omega_1=13298.2440$ and $\Omega_2$=2.3034.
\begin{table}[!ht]
\begin{center}
\caption{Some points belonging to the Pareto curve obtained by proposed multi-objective optimization problem ($t_f$=70 Days).}\label{tabpareto}
{\setlength{\tabulinesep}{1.2mm}
\begin{tabular}{ccccccc}
\hline
\ Point & \ $\Omega_1$ (Number of Individuals $\times$ Days) & \ $\Omega_2$ & \ $S(t_f)$  & \ $I(t_f)$ & \ $R(t_f)$ & \ $W(t_f)$ \\ \hline
\  A  & \  8963.7775   & \  6.9358      &  \ 135.2256  & \  2.1653      & \   769.0921 & \     141698.2905      \\
\  B  & \  56644.0350   & \  1.2940   &  \  33749.2312 & \  907.8714      & \ 4860.0582  & \   103087.6259       \\
\  C  & \  13298.2440   & \ 2.3034   &  \ 13697.0991  & \  20.4643      & \ 1140.9894  & \   127746.2276     \\\hline
\end{tabular}}
\end{center}
\end{table}

In Figures~\ref{Fig7newb} -- \ref{Fig7newf} are presented  the susceptible-infectious-removed populations profiles, control variable strategy and number of vaccinated individuals' profiles considering the points presented in Table~\ref{tabpareto}. In Figure \ref{Fig7newe} it is possible to observe the activation of the control variable when vaccine is introduced. Besides, in both results obtained, the action of such treatment is readily verified in the population during a larger interval of time in the beginning of the vaccine administration. In Figures~\ref{Fig7newb}, \ref{Fig7newc}, \ref{Fig7newd} and \ref{Fig7newf} the susceptible, infectious, recovered and number of vaccines profiles are presented, respectively, for each point described in Table~\ref{tabpareto}. In these figures we can visualize the importance of the control strategy used. For example, the points A and C are good choices in terms of the minimization of infected individuals, although the point A has a highest value in terms of the objective $\Omega_2$. On the other hand, point B is satisfactory in terms of minimizing the amount of vaccines administered, but, from a clinical point of view, it is not a good choice, as the number of infected individuals is not minimized.

\section{Conclusions} \label{eq:conclusions}

In this contribution it is proposed and solved an inverse problem to simulate the dynamic behavior of novel coronavirus disease (COVID-19) considering real data from China. The parameters of the compartmental SIR (Susceptible, Infectious and Recovered) model were determined by using Differential Evolution (DE). Considering the parameters obtained with the solution of the proposed inverse problem, two optimal control problems were proposed. The first consists on minimizing the quantity of infected individuals. In this case, an inequality that represents the quantity of vaccines available was analyzed. The second optimal control problem considers minimizing together the quantity of infected individuals and the prescribed vaccine during the treatment. This problem was solved using Multi-Objective Differential Evolution (MODE). In general, the solution of the proposed multi-objective optimal control problem provides information from which an optimal strategy for vaccine administration can be defined.

The use of mathematical models associated with optimization tools may contribute to decision making in situations of this type. It is important to emphasize that the quality of the results is dependent on the quality of the experimental data considered. In this context, one may cite the following limitations regarding the SIR model: $i$) poor quality of reported official data and; $ii$) the simplifications of the model, as for example terms as birth rate, differential vaccination rate, weather changes and its effect on the epidemiology. Finally, it is worth mentioning that the problem formulated in this work is not normally considered in the specialized literature (only the minimization of the infected individuals is normally proposed). In this context, the formulation of the multi-objective optimization problem and its solution by using MODE represents the main contribution of this work.



\section*{Acknowledgements}

\noindent This study was financed in part by the Coordena\c{c}{\~a}o de Aperfei\c{c}oamente de Pessoal de N{\'i}vel Superior---Brasil (CAPES)---Finance Code 001, Funda\c{c}{\~a}o Carlos Chagas Filho de Amparo {\`a} Pesquisa do Estado do Rio de Janeiro (FAPERJ), and Conselho Nacional de Desenvolvimento Cient{\'i}fico e Tecnol{\'o}gico (CNPq).

\bibliographystyle{elsarticle-harv} 
\bibliography{bibliography}

\begin{thebibliography}{51}
\expandafter\ifx\csname natexlab\endcsname\relax\def\natexlab#1{#1}\fi
\providecommand{\url}[1]{\texttt{#1}}
\providecommand{\href}[2]{#2}
\providecommand{\path}[1]{#1}
\providecommand{\DOIprefix}{doi:}
\providecommand{\ArXivprefix}{arXiv:}
\providecommand{\URLprefix}{URL: }
\providecommand{\Pubmedprefix}{pmid:}
\providecommand{\doi}[1]{\href{http://dx.doi.org/#1}{\path{#1}}}
\providecommand{\Pubmed}[1]{\href{pmid:#1}{\path{#1}}}
\providecommand{\bibinfo}[2]{#2}
\ifx\xfnm\relax \def\xfnm[#1]{\unskip,\space#1}\fi
\bibitem[{Al-Sheikh(2013)}]{AlSheikh2012}
\bibinfo{author}{Al-Sheikh, S.A.}, \bibinfo{year}{2013}.
\newblock \bibinfo{title}{{Modeling and Analysis of an SEIR Epidemic Model with
  a Limited Resource for Treatment}}.
\newblock \bibinfo{journal}{Global Journal of Science Frontier Research
  Mathematics and Decision Sciences} \bibinfo{volume}{12},
  \bibinfo{pages}{56--66}.
\bibitem[{Azam et~al.(2020)Azam, Mac{\'i}as-D{\'i}az, Ahmed, Khan, Iqbal,
  Rafiq, Nisar and Ahmad}]{ShumailaAzametal}
\bibinfo{author}{Azam, S.}, \bibinfo{author}{Mac{\'i}as-D{\'i}az, J.E.},
  \bibinfo{author}{Ahmed, N.}, \bibinfo{author}{Khan, I.},
  \bibinfo{author}{Iqbal, M.S.}, \bibinfo{author}{Rafiq, M.},
  \bibinfo{author}{Nisar, K.S.}, \bibinfo{author}{Ahmad, M.O.},
  \bibinfo{year}{2020}.
\newblock \bibinfo{title}{{Numerical Modeling and Theoretical Analysis of a
  Nonlinear Advection-reaction Epidemic System}}.
\newblock \bibinfo{journal}{Computer Methods and Programs in Biomedicine}
  \bibinfo{volume}{193}, \bibinfo{pages}{105429}.
\newblock \DOIprefix\doi{10.1016/j.cmpb.2020.105429}.
\bibitem[{Bauch et~al.(2009)Bauch, Szusz and Garrison}]{BauchSzuszGarrison2009}
\bibinfo{author}{Bauch, C.}, \bibinfo{author}{Szusz, E.},
  \bibinfo{author}{Garrison, L.}, \bibinfo{year}{2009}.
\newblock \bibinfo{title}{{Scheduling of Measles Vaccination in Low-income
  Countries: Projections of a Dynamic Model}}.
\newblock \bibinfo{journal}{Vaccine} \bibinfo{volume}{27},
  \bibinfo{pages}{4090--4098}.
\newblock \DOIprefix\doi{10.1016/j.vaccine.2009.04.079}.
\bibitem[{Benvenuto et~al.(2020)Benvenuto, Giovanetti, Vassallo, Angeletti and
  Ciccozzi}]{BenvenutoGiovanettiVassalloAngelettiCiccozzi2020}
\bibinfo{author}{Benvenuto, D.}, \bibinfo{author}{Giovanetti, M.},
  \bibinfo{author}{Vassallo, L.}, \bibinfo{author}{Angeletti, S.},
  \bibinfo{author}{Ciccozzi, M.}, \bibinfo{year}{2020}.
\newblock \bibinfo{title}{{Application of the ARIMA Model on the COVID-2019
  Epidemic Dataset}}.
\newblock \bibinfo{journal}{Data in Brief} \bibinfo{volume}{29},
  \bibinfo{pages}{105340}.
\newblock \DOIprefix\doi{10.1016/j.dib.2020.105340}.
\bibitem[{Biegler et~al.(2002)Biegler, Cervantes and
  W{\"a}chter}]{Biegleretal2002}
\bibinfo{author}{Biegler, L.T.}, \bibinfo{author}{Cervantes, A.M.},
  \bibinfo{author}{W{\"a}chter, A.}, \bibinfo{year}{2002}.
\newblock \bibinfo{title}{{Advances in Simultaneous Strategies for Dynamic
  Process Optimization}}.
\newblock \bibinfo{journal}{Chemical Engineering Science} \bibinfo{volume}{57},
  \bibinfo{pages}{575--593}.
\newblock \DOIprefix\doi{10.1016/S0009-2509(01)00376-1}.
\bibitem[{Biswas et~al.(2014)Biswas, Paiva and Pinho}]{BiswasPaivaPinho2014}
\bibinfo{author}{Biswas, M.H.A.}, \bibinfo{author}{Paiva, L.T.},
  \bibinfo{author}{Pinho, M.R.}, \bibinfo{year}{2014}.
\newblock \bibinfo{title}{{A SEIR Model for Control of Infectious Diseases with
  Constraints}}.
\newblock \bibinfo{journal}{Mathematical Biosciences and Engineering}
  \bibinfo{volume}{11}, \bibinfo{pages}{761}.
\newblock \DOIprefix\doi{10.3934/mbe.2014.11.761}.
\bibitem[{Blackwood and Childs(2018)}]{BlackwoodChilds2018}
\bibinfo{author}{Blackwood, J.C.}, \bibinfo{author}{Childs, L.M.},
  \bibinfo{year}{2018}.
\newblock \bibinfo{title}{{An Introduction to Compartmental Modeling for the
  Budding Infectious Disease Modeler}}.
\newblock \bibinfo{journal}{Letters in Biomathematics} \bibinfo{volume}{5},
  \bibinfo{pages}{195--221}.
\newblock \DOIprefix\doi{10.1080/23737867.2018.1509026}.
\bibitem[{Bowong and Kurths(2010)}]{BowongKurths2010}
\bibinfo{author}{Bowong, S.}, \bibinfo{author}{Kurths, J.},
  \bibinfo{year}{2010}.
\newblock \bibinfo{title}{{Parameter Estimation Based Synchronization for an
  Epidemic Model with Application to Tuberculosis in Cameroon}}.
\newblock \bibinfo{journal}{Physics Letters A} \bibinfo{volume}{374},
  \bibinfo{pages}{4496--4505}.
\newblock \DOIprefix\doi{10.1016/j.physleta.2010.09.008}.
\bibitem[{Bryson and Ho(1975)}]{BrysonHo1975}
\bibinfo{author}{Bryson, Jr, A.E.}, \bibinfo{author}{Ho, Y.C.},
  \bibinfo{year}{1975}.
\newblock \bibinfo{title}{{Applied Optimal Control: Optimization, Estimation
  and Control}}.
\newblock \bibinfo{publisher}{Taylor \& Francis}.
\bibitem[{Chan et~al.(2020)Chan, Yuan, Kok, To, Chu, Yang, Xing, Liu, Yip,
  Poon, Tsoi, Lo, Chan, Poon, Chan, Ip, Cai, Cheng, Chen, Hui and
  Yuen}]{ChanYuanKokToChu2020}
\bibinfo{author}{Chan, J.F.W.}, \bibinfo{author}{Yuan, S.},
  \bibinfo{author}{Kok, K.H.}, \bibinfo{author}{To, K.K.W.},
  \bibinfo{author}{Chu, H.}, \bibinfo{author}{Yang, J.}, \bibinfo{author}{Xing,
  F.}, \bibinfo{author}{Liu, J.}, \bibinfo{author}{Yip, C.C.Y.},
  \bibinfo{author}{Poon, R.W.S.}, \bibinfo{author}{Tsoi, H.W.},
  \bibinfo{author}{Lo, S.K.F.}, \bibinfo{author}{Chan, K.H.},
  \bibinfo{author}{Poon, V.K.M.}, \bibinfo{author}{Chan, W.M.},
  \bibinfo{author}{Ip, J.D.}, \bibinfo{author}{Cai, J.P.},
  \bibinfo{author}{Cheng, V.C.C.}, \bibinfo{author}{Chen, H.},
  \bibinfo{author}{Hui, C.K.M.}, \bibinfo{author}{Yuen, K.Y.},
  \bibinfo{year}{2020}.
\newblock \bibinfo{title}{{A Familial Cluster of Pneumonia Associated with the
  2019 Novel Coronavirus Indicating Person-to-person Transmission: a Study of a
  Family Cluster}}.
\newblock \bibinfo{journal}{The Lancet} \bibinfo{volume}{395},
  \bibinfo{pages}{514--523}.
\newblock \DOIprefix\doi{10.1016/s0140-6736(20)30154-9}.
\bibitem[{Cooper et~al.(2016)Cooper, Bastola, Gandhi, Ghersi, Hinrichs, Morien
  and Fruhling}]{Cooper2016}
\bibinfo{author}{Cooper, K.M.}, \bibinfo{author}{Bastola, D.R.},
  \bibinfo{author}{Gandhi, R.}, \bibinfo{author}{Ghersi, D.},
  \bibinfo{author}{Hinrichs, S.}, \bibinfo{author}{Morien, M.},
  \bibinfo{author}{Fruhling, A.}, \bibinfo{year}{2016}.
\newblock \bibinfo{title}{{Forecasting the Spread of Mosquito-Borne Disease
  using Publicly Accessible Data: A Case Study in Chikungunya}}.
\newblock \bibinfo{journal}{AMIA Annu Symp Proc} \bibinfo{volume}{2016},
  \bibinfo{pages}{431--440}.
\bibitem[{Deb(2001)}]{Deb}
\bibinfo{author}{Deb, K.}, \bibinfo{year}{2001}.
\newblock \bibinfo{title}{{Multi-objective Optimization Using Evolutionary
  Algorithms}}.
\newblock Wiley-Interscience Series in Systems and Optimization.
  \bibinfo{edition}{1} ed., \bibinfo{publisher}{John Wiley \& Sons}.
\bibitem[{Edgeworth(1881)}]{Edgeworth}
\bibinfo{author}{Edgeworth, F.Y.}, \bibinfo{year}{1881}.
\newblock \bibinfo{title}{Mathematical Psychics}.
\newblock \bibinfo{publisher}{Nabu Press}.
\bibitem[{Feehery and Barton(1996)}]{Feehery1998}
\bibinfo{author}{Feehery, W.F.}, \bibinfo{author}{Barton, P.I.},
  \bibinfo{year}{1996}.
\newblock \bibinfo{title}{{Dynamic Simulation and Optimization with Inequality
  Path Constraints}}.
\newblock \bibinfo{journal}{Computers {\&} Chemical Engineering}
  \bibinfo{volume}{20}, \bibinfo{pages}{S707--S712}.
\newblock \DOIprefix\doi{10.1016/0098-1354(96)00127-5}.
\bibitem[{Forgoston and Schwartz(2013)}]{Forgoston2013}
\bibinfo{author}{Forgoston, E.}, \bibinfo{author}{Schwartz, I.B.},
  \bibinfo{year}{2013}.
\newblock \bibinfo{title}{{Predicting Unobserved Exposures from Seasonal
  Epidemic Data}}.
\newblock \bibinfo{journal}{Bulletin of Mathematical Biology}
  \bibinfo{volume}{75}, \bibinfo{pages}{1450--1471}.
\newblock \DOIprefix\doi{10.1007/s11538-013-9855-0}.
\bibitem[{Gorbalenya et~al.(2020)Gorbalenya, Baker, Baric, de, Drosten,
  Gulyaeva, Haagmans, Lauber, Leontovich, Neuman, Penzar, Perlman, Poon,
  Samborskiy, Sidorov, Sola and Ziebuhr}]{GorbalenyaGulyaeva2020}
\bibinfo{author}{Gorbalenya, A.E.}, \bibinfo{author}{Baker, S.C.},
  \bibinfo{author}{Baric, R.S.}, \bibinfo{author}{de, Groot, R.J.},
  \bibinfo{author}{Drosten, C.}, \bibinfo{author}{Gulyaeva, A.A.},
  \bibinfo{author}{Haagmans, B.L.}, \bibinfo{author}{Lauber, C.},
  \bibinfo{author}{Leontovich, A.M.}, \bibinfo{author}{Neuman, B.W.},
  \bibinfo{author}{Penzar, D.}, \bibinfo{author}{Perlman, S.},
  \bibinfo{author}{Poon, L.L.}, \bibinfo{author}{Samborskiy, D.},
  \bibinfo{author}{Sidorov, I.A.}, \bibinfo{author}{Sola, I.},
  \bibinfo{author}{Ziebuhr, J.}, \bibinfo{year}{2020}.
\newblock \bibinfo{title}{{Severe Acute Respiratory Syndrome-related
  Coronavirus: the Species and Its Viruses---A Statement of the Coronavirus
  Study Group}}.
\newblock \bibinfo{journal}{bioRxiv} \DOIprefix\doi{10.1101/2020.02.07.937862}.
\bibitem[{Hethcote(2000)}]{Hethcote2000}
\bibinfo{author}{Hethcote, H.W.}, \bibinfo{year}{2000}.
\newblock \bibinfo{title}{The mathematics of infectious diseases}.
\newblock \bibinfo{journal}{{SIAM} Review} \bibinfo{volume}{42},
  \bibinfo{pages}{599--653}.
\newblock \DOIprefix\doi{10.1137/s0036144500371907}.
\bibitem[{Hu et~al.(2005)Hu, Coello and Huang}]{Hu}
\bibinfo{author}{Hu, X.}, \bibinfo{author}{Coello, Coello, C.A.},
  \bibinfo{author}{Huang, Z.}, \bibinfo{year}{2005}.
\newblock \bibinfo{title}{{A New Multi-objective Evolutionary Algorithm:
  Neighbourhood Exploring Evolution Strategy}}.
\newblock \bibinfo{journal}{Engineering Optimization} \bibinfo{volume}{37},
  \bibinfo{pages}{351--379}.
\newblock \DOIprefix\doi{10.1080/03052150500035658}.
\bibitem[{Huang et~al.(2020)Huang, Wang, Li, Ren, Zhao, Hu, Zhang, Fan, Xu, Gu,
  Cheng, Yu, Xia, Wei, Wu, Xie, Yin, Li, Liu, Xiao, Gao, Guo, Xie, Wang, Jiang,
  Gao, Jin, Wang and Cao}]{HuangWangLiRenZhaoHuZhangFanXuGu2020}
\bibinfo{author}{Huang, C.}, \bibinfo{author}{Wang, Y.}, \bibinfo{author}{Li,
  X.}, \bibinfo{author}{Ren, L.}, \bibinfo{author}{Zhao, J.},
  \bibinfo{author}{Hu, Y.}, \bibinfo{author}{Zhang, L.}, \bibinfo{author}{Fan,
  G.}, \bibinfo{author}{Xu, J.}, \bibinfo{author}{Gu, X.},
  \bibinfo{author}{Cheng, Z.}, \bibinfo{author}{Yu, T.}, \bibinfo{author}{Xia,
  J.}, \bibinfo{author}{Wei, Y.}, \bibinfo{author}{Wu, W.},
  \bibinfo{author}{Xie, X.}, \bibinfo{author}{Yin, W.}, \bibinfo{author}{Li,
  H.}, \bibinfo{author}{Liu, M.}, \bibinfo{author}{Xiao, Y.},
  \bibinfo{author}{Gao, H.}, \bibinfo{author}{Guo, L.}, \bibinfo{author}{Xie,
  J.}, \bibinfo{author}{Wang, G.}, \bibinfo{author}{Jiang, R.},
  \bibinfo{author}{Gao, Z.}, \bibinfo{author}{Jin, Q.}, \bibinfo{author}{Wang,
  J.}, \bibinfo{author}{Cao, B.}, \bibinfo{year}{2020}.
\newblock \bibinfo{title}{{Clinical Features of Patients Infected with 2019
  Novel Coronavirus in Wuhan, China}}.
\newblock \bibinfo{journal}{The Lancet} \bibinfo{volume}{395},
  \bibinfo{pages}{497--506}.
\newblock \DOIprefix\doi{10.1016/s0140-6736(20)30183-5}.
\bibitem[{{Johns Hopkins Resource Center}(2020 (accessed April 03,
  2020))}]{JHDC}
\bibinfo{author}{{Johns Hopkins Resource Center}}, \bibinfo{year}{2020
  (accessed April 03, 2020)}.
\newblock \bibinfo{title}{Mapping 2019-nCoV}.
\newblock
  \bibinfo{note}{\url{https://systems.jhu.edu/research/public-health/ncov/}}.
\bibitem[{Keeling and Rohani(2007)}]{KeelingRohani2011}
\bibinfo{author}{Keeling, M.J.}, \bibinfo{author}{Rohani, P.},
  \bibinfo{year}{2007}.
\newblock \bibinfo{title}{{Modeling Infectious Diseases in Humans and
  Animals}}.
\newblock \bibinfo{publisher}{Princeton University Press}.
\bibitem[{Kim et~al.(2019)Kim, Byun and Jung}]{KimByunJung2019}
\bibinfo{author}{Kim, S.}, \bibinfo{author}{Byun, J.H.}, \bibinfo{author}{Jung,
  I.H.}, \bibinfo{year}{2019}.
\newblock \bibinfo{title}{{Global Stability of an SEIR Epidemic Model Where
  Empirical Distribution of Incubation Period is Approximated by Coxian
  Distribution}}.
\newblock \bibinfo{journal}{Advances in Difference Equations}
  \DOIprefix\doi{10.1186/s13662-019-2405-9}.
\bibitem[{Li and Cui(2013)}]{LiCui2013}
\bibinfo{author}{Li, J.}, \bibinfo{author}{Cui, N.}, \bibinfo{year}{2013}.
\newblock \bibinfo{title}{{Dynamic Analysis of an {SEIR} Model with Distinct
  Incidence for Exposed and Infectives}}.
\newblock \bibinfo{journal}{The Scientific World Journal}
  \bibinfo{volume}{2013}, \bibinfo{pages}{1--5}.
\newblock \DOIprefix\doi{10.1155/2013/871393}.
\bibitem[{Li and Muldowney(1995)}]{LiMuldowney1995}
\bibinfo{author}{Li, M.Y.}, \bibinfo{author}{Muldowney, J.S.},
  \bibinfo{year}{1995}.
\newblock \bibinfo{title}{{Global Stability for the {SEIR} Model in
  Epidemiology}}.
\newblock \bibinfo{journal}{Mathematical Biosciences} \bibinfo{volume}{125},
  \bibinfo{pages}{155--164}.
\newblock \DOIprefix\doi{10.1016/0025-5564(95)92756-5}.
\bibitem[{Li et~al.(2020)Li, Guan, Wu, Wang, Zhou, Tong, Ren, Leung, Lau, Wong,
  Xing, Xiang, Wu, Li, Chen, Li, Liu, Zhao, Liu, Tu, Chen, Jin, Yang, Wang,
  Zhou, Wang, Liu, Luo, Liu, Shao, Li, Tao, Yang, Deng, Liu, Ma, Zhang, Shi,
  Lam, Wu, Gao, Cowling, Yang, Leung and Feng}]{LiGuanWu2020}
\bibinfo{author}{Li, Q.}, \bibinfo{author}{Guan, X.}, \bibinfo{author}{Wu, P.},
  \bibinfo{author}{Wang, X.}, \bibinfo{author}{Zhou, L.},
  \bibinfo{author}{Tong, Y.}, \bibinfo{author}{Ren, R.},
  \bibinfo{author}{Leung, K.S.}, \bibinfo{author}{Lau, E.H.},
  \bibinfo{author}{Wong, J.Y.}, \bibinfo{author}{Xing, X.},
  \bibinfo{author}{Xiang, N.}, \bibinfo{author}{Wu, Y.}, \bibinfo{author}{Li,
  C.}, \bibinfo{author}{Chen, Q.}, \bibinfo{author}{Li, D.},
  \bibinfo{author}{Liu, T.}, \bibinfo{author}{Zhao, J.}, \bibinfo{author}{Liu,
  M.}, \bibinfo{author}{Tu, W.}, \bibinfo{author}{Chen, C.},
  \bibinfo{author}{Jin, L.}, \bibinfo{author}{Yang, R.}, \bibinfo{author}{Wang,
  Q.}, \bibinfo{author}{Zhou, S.}, \bibinfo{author}{Wang, R.},
  \bibinfo{author}{Liu, H.}, \bibinfo{author}{Luo, Y.}, \bibinfo{author}{Liu,
  Y.}, \bibinfo{author}{Shao, G.}, \bibinfo{author}{Li, H.},
  \bibinfo{author}{Tao, Z.}, \bibinfo{author}{Yang, Y.}, \bibinfo{author}{Deng,
  Z.}, \bibinfo{author}{Liu, B.}, \bibinfo{author}{Ma, Z.},
  \bibinfo{author}{Zhang, Y.}, \bibinfo{author}{Shi, G.}, \bibinfo{author}{Lam,
  T.T.}, \bibinfo{author}{Wu, J.T.}, \bibinfo{author}{Gao, G.F.},
  \bibinfo{author}{Cowling, B.J.}, \bibinfo{author}{Yang, B.},
  \bibinfo{author}{Leung, G.M.}, \bibinfo{author}{Feng, Z.},
  \bibinfo{year}{2020}.
\newblock \bibinfo{title}{{Early Transmission Dynamics in Wuhan, China, of
  Novel Coronavirus–Infected Pneumonia}}.
\newblock \bibinfo{journal}{New England Journal of Medicine}
  \bibinfo{volume}{382}, \bibinfo{pages}{1199--1207}.
\newblock \DOIprefix\doi{10.1056/NEJMoa2001316}.
\bibitem[{Lin et~al.(2020)Lin, Zhao, Gao, Lou, Yang, Musa, Wang, Cai, Wang,
  Yang and He}]{LinZhaoGaoLouYangMusaWangCaiWangYangHe2019}
\bibinfo{author}{Lin, Q.}, \bibinfo{author}{Zhao, S.}, \bibinfo{author}{Gao,
  D.}, \bibinfo{author}{Lou, Y.}, \bibinfo{author}{Yang, S.},
  \bibinfo{author}{Musa, S.S.}, \bibinfo{author}{Wang, M.H.},
  \bibinfo{author}{Cai, Y.}, \bibinfo{author}{Wang, W.}, \bibinfo{author}{Yang,
  L.}, \bibinfo{author}{He, D.}, \bibinfo{year}{2020}.
\newblock \bibinfo{title}{{A Conceptual Model for the Coronavirus Disease 2019
  (COVID-19) Outbreak in Wuhan, China with Individual Reaction and Governmental
  Action}}.
\newblock \bibinfo{journal}{International Journal of Infectious Diseases}
  \bibinfo{volume}{93}, \bibinfo{pages}{211--216}.
\newblock \DOIprefix\doi{=10.1016/j.ijid.2020.02.058}.
\bibitem[{Lobato(2004)}]{Lobato2004}
\bibinfo{author}{Lobato, F.S.}, \bibinfo{year}{2004}.
\newblock \bibinfo{title}{Hybrid Approach for Dynamic Optimization Problems}.
\newblock Master's thesis. Federal University of Uberl{\^a}ndia.
  \bibinfo{address}{Uberl{\^a}ndia}.
\newblock \bibinfo{note}{In Portuguese}.
\bibitem[{Lobato(2008)}]{Lobato2008}
\bibinfo{author}{Lobato, F.S.}, \bibinfo{year}{2008}.
\newblock \bibinfo{title}{Multi-objective Optimization for Engineering System
  Design}.
\newblock Ph.D. thesis. Federal University of Uberl{\^a}ndia.
  \bibinfo{address}{Uberl{\^a}ndia}.
\newblock \bibinfo{note}{In Portuguese}.
\bibitem[{Lobato et~al.(2016)Lobato, Machado and
  Steffen}]{LobatoSilverioSteffen2016}
\bibinfo{author}{Lobato, F.S.}, \bibinfo{author}{Machado, V.S.},
  \bibinfo{author}{Steffen, Jr, V.S.}, \bibinfo{year}{2016}.
\newblock \bibinfo{title}{{Determination of an Optimal Control Strategy for
  Drug Administration in Tumor Treatment Using Multi-objective Optimization
  Differential Evolution}}.
\newblock \bibinfo{journal}{Computer Methods and Programs in Biomedicine}
  \bibinfo{volume}{131}, \bibinfo{pages}{51--61}.
\newblock \DOIprefix\doi{10.1016/j.cmpb.2016.04.004}.
\bibitem[{Lobato and Steffen(2011)}]{LobatoSteffenJr}
\bibinfo{author}{Lobato, F.S.}, \bibinfo{author}{Steffen, Jr, V.},
  \bibinfo{year}{2011}.
\newblock \bibinfo{title}{{A New Multi-objective Optimization Algorithm Based
  on Differential Evolution and Neighborhood Exploring Evolution Strategy}}.
\newblock \bibinfo{journal}{Journal of Artificial Intelligence and Soft
  Computing Research} \bibinfo{volume}{1}, \bibinfo{pages}{259--267}.
\bibitem[{Lurie et~al.(2020)Lurie, Saville, Hatchett and Halton}]{Lurie2020}
\bibinfo{author}{Lurie, N.}, \bibinfo{author}{Saville, M.},
  \bibinfo{author}{Hatchett, R.}, \bibinfo{author}{Halton, J.},
  \bibinfo{year}{2020}.
\newblock \bibinfo{title}{{Developing Covid-19 Vaccines at Pandemic Speed}}.
\newblock \bibinfo{journal}{New England Journal of Medicine}
  \DOIprefix\doi{10.1056/NEJMp2005630}.
\bibitem[{Moura~Neto and Silva~Neto(2013)}]{MouraNetoSilvaNeto}
\bibinfo{author}{Moura~Neto, F.D.}, \bibinfo{author}{Silva~Neto, A.J.},
  \bibinfo{year}{2013}.
\newblock \bibinfo{title}{{An Introduction to Inverse Problems with
  Applications}}.
\newblock \bibinfo{publisher}{Springer Berlin Heidelberg},
  \bibinfo{address}{Berlin, Heidelberg}.
\newblock \DOIprefix\doi{10.1007/978-3-642-32557-1}.
\bibitem[{Mukandavire et~al.(2009)Mukandavire, Chiyaka, Garira and
  Musuka}]{MukandavireChiyakaGariraMusuka2009}
\bibinfo{author}{Mukandavire, Z.}, \bibinfo{author}{Chiyaka, C.},
  \bibinfo{author}{Garira, W.}, \bibinfo{author}{Musuka, G.},
  \bibinfo{year}{2009}.
\newblock \bibinfo{title}{{Mathematical Analysis of a Sex-structured HIV/AIDS
  Model with a Discrete Time Delay}}.
\newblock \bibinfo{journal}{Nonlinear Analysis: Theory, Methods \&
  Applications} \bibinfo{volume}{71}, \bibinfo{pages}{1082--1093}.
\newblock \DOIprefix\doi{10.1016/j.na.2008.11.026}.
\bibitem[{Neilan and Lenhart(2010)}]{NeilanLenhart2010}
\bibinfo{author}{Neilan, R.M.}, \bibinfo{author}{Lenhart, S.},
  \bibinfo{year}{2010}.
\newblock \bibinfo{title}{An introduction to optimal control with an
  application in disease modeling}, in: \bibinfo{editor}{Gumel, A.B.},
  \bibinfo{editor}{Lenhart, S.} (Eds.), \bibinfo{booktitle}{Modeling Paradigms
  and Analysis of Disease Trasmission Models}, \bibinfo{publisher}{American
  Mathematical Society}. pp. \bibinfo{pages}{67--81}.
\bibitem[{Pareto(1896)}]{Pareto}
\bibinfo{author}{Pareto, V.}, \bibinfo{year}{1896}.
\newblock \bibinfo{title}{Cours d'{\'E}conomie Politique}.
\newblock \bibinfo{publisher}{F. Rouge}, \bibinfo{address}{Lausanne}.
\bibitem[{Pesco et~al.(2014)Pesco, Bergero, Fabricius and Hozbor}]{Pesco2014}
\bibinfo{author}{Pesco, P.}, \bibinfo{author}{Bergero, P.},
  \bibinfo{author}{Fabricius, G.}, \bibinfo{author}{Hozbor, D.},
  \bibinfo{year}{2014}.
\newblock \bibinfo{title}{{Modelling the Effect of Changes in Vaccine
  Effectiveness and Transmission Contact Rates on Pertussis Epidemiology}}.
\newblock \bibinfo{journal}{Epidemics} \bibinfo{volume}{7},
  \bibinfo{pages}{13--21}.
\newblock \DOIprefix\doi{10.1016/j.epidem.2014.04.001}.
\bibitem[{Prem et~al.(2020)Prem, Liu, Russell, Kucharski, Eggo, Davies, Jit,
  Klepac, Flasche, Clifford, Pearson, Munday, Abbott, Gibbs, Rosello, Quilty,
  Jombart, Sun, Diamond, Gimma, van Zandvoort, Funk, Jarvis, Edmunds, Bosse and
  Hellewell}]{PremLiuRussellKucharskiEggo2020}
\bibinfo{author}{Prem, K.}, \bibinfo{author}{Liu, Y.},
  \bibinfo{author}{Russell, T.W.}, \bibinfo{author}{Kucharski, A.J.},
  \bibinfo{author}{Eggo, R.M.}, \bibinfo{author}{Davies, N.},
  \bibinfo{author}{Jit, M.}, \bibinfo{author}{Klepac, P.},
  \bibinfo{author}{Flasche, S.}, \bibinfo{author}{Clifford, S.},
  \bibinfo{author}{Pearson, C.A.B.}, \bibinfo{author}{Munday, J.D.},
  \bibinfo{author}{Abbott, S.}, \bibinfo{author}{Gibbs, H.},
  \bibinfo{author}{Rosello, A.}, \bibinfo{author}{Quilty, B.J.},
  \bibinfo{author}{Jombart, T.}, \bibinfo{author}{Sun, F.},
  \bibinfo{author}{Diamond, C.}, \bibinfo{author}{Gimma, A.},
  \bibinfo{author}{van Zandvoort, K.}, \bibinfo{author}{Funk, S.},
  \bibinfo{author}{Jarvis, C.I.}, \bibinfo{author}{Edmunds, W.J.},
  \bibinfo{author}{Bosse, N.I.}, \bibinfo{author}{Hellewell, J.},
  \bibinfo{year}{2020}.
\newblock \bibinfo{title}{{The Effect of Control Strategies to Reduce Social
  Mixing on Outcomes of the {COVID}-19 Epidemic in Wuhan, China: a Modelling
  Study}}.
\newblock \bibinfo{journal}{The Lancet Public Health}
  \DOIprefix\doi{10.1016/s2468-2667(20)30073-6}.
\bibitem[{Price et~al.(2005)Price, Storn and Lampinen}]{StornPriceLampinen}
\bibinfo{author}{Price, K.V.}, \bibinfo{author}{Storn, R.M.},
  \bibinfo{author}{Lampinen, J.A.}, \bibinfo{year}{2005}.
\newblock \bibinfo{title}{{Differential Evolution: A Practical Approach to
  Global Optimization}}.
\newblock \bibinfo{publisher}{Springer Berlin Heidelberg},
  \bibinfo{address}{Berlin, Heidelberg}.
\bibitem[{Riou and Althaus(2020)}]{RiouAlthaus2020}
\bibinfo{author}{Riou, J.}, \bibinfo{author}{Althaus, C.L.},
  \bibinfo{year}{2020}.
\newblock \bibinfo{title}{{Pattern of Early Human-to-human Transmission of
  Wuhan 2019 Novel Coronavirus (2019-nCoV), December 2019 to January 2020}}.
\newblock \bibinfo{journal}{Eurosurveillance} \bibinfo{volume}{25},
  \bibinfo{pages}{5}.
\newblock \DOIprefix\doi{10.2807/1560-7917.ES.2020.25.4.2000058}.
\bibitem[{Roda et~al.(2020)Roda, Varughese, Han and
  Li}]{RodaVarugheseHanLi2020}
\bibinfo{author}{Roda, W.C.}, \bibinfo{author}{Varughese, M.B.},
  \bibinfo{author}{Han, D.}, \bibinfo{author}{Li, M.Y.}, \bibinfo{year}{2020}.
\newblock \bibinfo{title}{{Why is it Difficult to Accurately Predict the
  {COVID}-19 Epidemic?}}
\newblock \bibinfo{journal}{Infectious Disease Modelling} \bibinfo{volume}{5},
  \bibinfo{pages}{271--281}.
\newblock \DOIprefix\doi{10.1016/j.idm.2020.03.001}.
\bibitem[{Shaman et~al.(2014)Shaman, Yang and Kandula}]{Shaman2014}
\bibinfo{author}{Shaman, J.}, \bibinfo{author}{Yang, W.},
  \bibinfo{author}{Kandula, S.}, \bibinfo{year}{2014}.
\newblock \bibinfo{title}{{Inference and Forecast of the Current West African
  Ebola Outbreak in Guinea, Sierra Leone and Liberia}}.
\newblock \bibinfo{journal}{PLoS Currents}
  \DOIprefix\doi{10.1371/currents.outbreaks.3408774290b1a0f2dd7cae877c8b8ff6}.
\bibitem[{Singh et~al.(2017)Singh, Srivastava and
  Arora}]{SinghSrivastavaArora2017}
\bibinfo{author}{Singh, P.}, \bibinfo{author}{Srivastava, S.K.},
  \bibinfo{author}{Arora, U.}, \bibinfo{year}{2017}.
\newblock \bibinfo{title}{{Stability of SEIR Model of Infectious Diseases with
  Human Immunity}}.
\newblock \bibinfo{journal}{Global Journal of Pure and Applied Mathematics}
  \bibinfo{volume}{13}, \bibinfo{pages}{1811--1819}.
\bibitem[{Storn and Price(1997)}]{StornPrice1995}
\bibinfo{author}{Storn, R.}, \bibinfo{author}{Price, K.}, \bibinfo{year}{1997}.
\newblock \bibinfo{title}{{Differential Evolution---A Simple and Efficient
  Heuristic for Global Optimization over Continuous Spaces}}.
\newblock \bibinfo{journal}{Journal of Global Optimization}
  \bibinfo{volume}{11}, \bibinfo{pages}{341--359}.
\newblock \DOIprefix\doi{10.1023/A:1008202821328}.
\bibitem[{Trawicki(2017)}]{Trawicki2017}
\bibinfo{author}{Trawicki, M.}, \bibinfo{year}{2017}.
\newblock \bibinfo{title}{{Deterministic SEIRs Epidemic Model for Modeling
  Vital Dynamics, Vaccinations, and Temporary Immunity}}.
\newblock \bibinfo{journal}{Mathematics} \bibinfo{volume}{5},
  \bibinfo{pages}{7}.
\newblock \DOIprefix\doi{10.3390/math5010007}.
\bibitem[{Wang et~al.(2020)Wang, Hu, Hu, Zhu, Liu, Zhang, Wang, Xiang, Cheng,
  Xiong, Zhao, Li, Wang and Peng}]{WangHuHu2020}
\bibinfo{author}{Wang, D.}, \bibinfo{author}{Hu, B.}, \bibinfo{author}{Hu, C.},
  \bibinfo{author}{Zhu, F.}, \bibinfo{author}{Liu, X.}, \bibinfo{author}{Zhang,
  J.}, \bibinfo{author}{Wang, B.}, \bibinfo{author}{Xiang, H.},
  \bibinfo{author}{Cheng, Z.}, \bibinfo{author}{Xiong, Y.},
  \bibinfo{author}{Zhao, Y.}, \bibinfo{author}{Li, Y.}, \bibinfo{author}{Wang,
  X.}, \bibinfo{author}{Peng, Z.}, \bibinfo{year}{2020}.
\newblock \bibinfo{title}{{Clinical Characteristics of 138 Hospitalized
  Patients With 2019 Novel Coronavirus–Infected Pneumonia in Wuhan, China}}.
\newblock \bibinfo{journal}{JAMA} \bibinfo{volume}{323},
  \bibinfo{pages}{1061--1069}.
\newblock \DOIprefix\doi{10.1001/jama.2020.1585}.
\bibitem[{Wei et~al.(2020)Wei, Zheng, Lei, Wu, Verma, Liu, Wei, Bi, Hu and
  Han}]{WeiZhengLeiWuVermaLiuWeiBiHuHan2020}
\bibinfo{author}{Wei, W.}, \bibinfo{author}{Zheng, D.}, \bibinfo{author}{Lei,
  Y.}, \bibinfo{author}{Wu, S.}, \bibinfo{author}{Verma, V.},
  \bibinfo{author}{Liu, Y.}, \bibinfo{author}{Wei, X.}, \bibinfo{author}{Bi,
  J.}, \bibinfo{author}{Hu, D.}, \bibinfo{author}{Han, G.},
  \bibinfo{year}{2020}.
\newblock \bibinfo{title}{{Radiotherapy Workflow and Protection Procedures
  During the Coronavirus Disease 2019 (COVID-19) Outbreak: Experience of the
  Hubei Cancer Hospital in Wuhan, China}}.
\newblock \bibinfo{journal}{Radiotherapy and Oncology}
  \DOIprefix\doi{10.1016/j.radonc.2020.03.029}.
\bibitem[{Weiss(2013)}]{Weiss2013}
\bibinfo{author}{Weiss, H.H.}, \bibinfo{year}{2013}.
\newblock \bibinfo{title}{{The SIR Model and the Foundations of Public
  Health}}.
\newblock \bibinfo{journal}{Publicaci{\'o} electr{\`o}nica de divulgaci{\'o}
  del Departament de Matem{\`a}tiques de la Universitat Aut{\`o}noma de
  Barcelona} \bibinfo{volume}{2013}, \bibinfo{pages}{17}.
\bibitem[{Widyaningsih et~al.(2018)Widyaningsih, Saputro and
  Nugroho}]{Widyaningsih2018}
\bibinfo{author}{Widyaningsih, P.}, \bibinfo{author}{Saputro, D.R.S.},
  \bibinfo{author}{Nugroho, A.W.}, \bibinfo{year}{2018}.
\newblock \bibinfo{title}{{Susceptible Exposed Infected Recovery ({SEIR}) Model
  with Immigration: Equilibria Points and its Application}}.
\newblock \bibinfo{journal}{AIP Conference Proceedings} \bibinfo{volume}{2014},
  \bibinfo{pages}{020165}.
\newblock \DOIprefix\doi{10.1063/1.5054569}.
\bibitem[{{World Health Organization}(2020 (accessed April 8,
  2020))}]{WorldHealthOrganization}
\bibinfo{author}{{World Health Organization}}, \bibinfo{year}{2020 (accessed
  April 8, 2020)}.
\newblock \bibinfo{title}{{Naming the Coronavirus Disease (COVID-19) and the
  Virus that Causes it}}.
\newblock
  \bibinfo{note}{\url{https://www.who.int/emergencies/diseases/novel-coronavirus-2019/technical-guidance/naming-the-coronavirus-disease-(covid-2019)-and-the-virus-that-causes-it}}.
\bibitem[{Zhai et~al.(2020)Zhai, Ding, Wu, Long, Zhong and
  Li}]{PanZhaiYanbingDingXiaWuJunkeLongYanjunZhongYimingLi2020}
\bibinfo{author}{Zhai, P.}, \bibinfo{author}{Ding, Y.}, \bibinfo{author}{Wu,
  X.}, \bibinfo{author}{Long, J.}, \bibinfo{author}{Zhong, Y.},
  \bibinfo{author}{Li, Y.}, \bibinfo{year}{2020}.
\newblock \bibinfo{title}{{The Epidemiology, Diagnosis and Treatment of
  COVID-19}}.
\newblock \bibinfo{journal}{International Journal of Antimicrobial Agents} ,
  \bibinfo{pages}{105955}\DOIprefix\doi{10.1016/j.ijantimicag.2020.105955}.
\bibitem[{Zitzler and Thiele(1999)}]{Zitzler}
\bibinfo{author}{Zitzler, E.}, \bibinfo{author}{Thiele, L.},
  \bibinfo{year}{1999}.
\newblock \bibinfo{title}{{Multiobjective Evolutionary Algorithms: a
  Comparative Case Study and the Strength Pareto Approach}}.
\newblock \bibinfo{journal}{IEEE Transactions on Evolutionary Computation}
  \bibinfo{volume}{3}, \bibinfo{pages}{257--271}.
\newblock \DOIprefix\doi{10.1109/4235.797969}.

\end{thebibliography}
\end{document}